\documentclass[]{pasj01}

\usepackage{natbib}
\usepackage{booktabs}
\bibpunct{(}{)}{;}{a}{}{,}

\usepackage{color}
\usepackage{ulem}

\begin{document} 
\Received{2018/08/13}
\Accepted{2019/01/15}

\title{A molecular line survey toward the nearby
galaxies NGC 1068, NGC 253, and IC 342 at 3 mm
with the Nobeyama 45 m radio telescope:
The data \\
}

\author{Shuro \textsc{Takano}\altaffilmark{1,2,*}}%
\altaffiltext{1}{Nobeyama Radio Observatory,
National Astronomical Observatory of Japan,
462-2 Nobeyama, Minamimaki, Minamisaku,
Nagano 384-1305}
\altaffiltext{2}{Department of Astronomical Science,
SOKENDAI (The Graduate University for Advanced Studies),
462-2 Nobeyama, Minamimaki, Minamisaku,
Nagano 384-1305}
\email{takano.shuro@nihon-u.ac.jp}

\author{Taku \textsc{Nakajima},\altaffilmark{3}}
\altaffiltext{3}{Institute for Space-Earth Environmental Research,
Nagoya University,
Furo-cho, Chikusa-ku, 
Nagoya, Aichi 464-8601}

\author{Kotaro \textsc{Kohno}\altaffilmark{4,5}}
\altaffiltext{4}{Institute of Astronomy, 
Graduate School of Science,
The University of Tokyo,
2-21-1 Osawa, Mitaka, Tokyo 181-0015}
\altaffiltext{5}{Research Center for the Early Universe, 
Graduate School of Science,
The University of Tokyo, 
7-3-1 Hongo, Bunkyo-ku, Tokyo 113-0033}
\altaffiltext{*}{Present address: Department of Physics, General Studies, College of
Engineering, Nihon University, Tamuramachi, Koriyama, Fukushima 963-8642}


\maketitle

\begin{abstract}
We present observational data of a molecular line survey
toward the nearby
galaxies NGC 1068, NGC 253, and IC 342 at the wavelengths
of 3 mm ($\sim$85--116 GHz) obtained
with the Nobeyama 45 m radio telescope. 
In IC 342 the line survey with high spectral resolution
at the 3 mm region was reported for the first time.
NGC 1068 is a nearby gas-rich 
galaxy with X-rays from an active galactic nucleus 
(AGN), and NGC 253 and IC 342 are nearby gas-rich 
galaxies with prototypical starbursts.  
These galaxies 
are useful to study the impacts of X-rays and ultraviolet radiation 
on molecular abundances.  
The survey was carried out with the resulting rms noise 
level of a few mK ($T\rm{_A^*}$).  
As a result we could obtain almost complete data of these 
galaxies at the 3 mm region: 
We detected 19--23 molecular species depending on 
the galaxies including several new 
detections (e.g., cyclic-C$_3$H$_2$ in IC 342).  
We found that the intensities of HCN, 
CN, and HC$_3$N relative to $^{13}$CO are significantly strong in NGC 1068 
compared to those in NGC 253 and IC 342.   
On the other hand, CH$_3$CCH was not detected in NGC 1068. 
We obtained these results with the narrow beam  (15$''$.2--19$''$.1) 
of the 45 m telescope
among the single-dish telescopes, and in particular 
selectively observed the molecular gas close to the 
circumnuclear disk (CND) in NGC 1068.
Our line intensities in NGC 1068 were compared to 
those obtained with the IRAM 30 m
radio telescope already reported.
As a result,
the intensity ratio of each
line was found to have information on the spatial distribution.
Our observations obtained the line intensities and 
stringent constraints
on the upper limit for the three galaxies with such narrow beam, and
consequently, 
the data will be a basis for further observations
with high spatial resolution.
\end{abstract}

\section{Introduction}

Recent progress of radio telescopes and related instruments
enabled us observations 
with high sensitivity and with wide-band spectroscopy.
The progress of receivers, intermediate frequency (IF) systems,
analog-to-digital (AD) converters, and
digital spectrometers achieved, for example, simultaneous  
observations of  
$\sim$ 16 GHz bandwidth in total for upper and lower sidebands.
As a result, our knowledge of interstellar molecules and atoms
are drastically increasing. 
Spectral lines of them are indispensable as probes of 
astrophysical phenomena, in particular, 
deep inside of dust obscured regions, and 
such lines are also indispensable for astrochemical and 
astrobiological studies.

So far $\sim$200 molecular species have been detected in interstellar
space and circumstellar envelopes \citep[e.g., The Cologne Database for 
Molecular Spectroscopy: CDMS,  ][]{muelle2001, muelle2005, endres2016}. 
Significant fraction of them has also been detected in
external galaxies.
Galaxies show a wide range of
environments such as active galactic
nuclei (AGNs), starbursts, arm-interarms,
bars, mergers, different metallicity and so on.
Effects of such environments on molecular and atomic gas are
quite interesting to probe such environments themselves and to study
their chemistry, which seems to be different from those in
quiescent Galactic objects.  
In particular, AGNs and starbursts are quite energetic phenomena,
which do not exist in our Galaxy.
The effects of such energetic environments on 
molecules and atoms are 
important topics to search for good probes of AGNs and/or starbursts, 
and to study molecular and atomic processes with X-rays and ultraviolet
radiation.
Many observational studies of such environments have already been 
reported \citep[e.g.,][]{jackso1993, kohno1996, kohno2003, usero2004, werf2010, 
rangwa2011, nakaji2011, 
aladro2013, izumi2013, garcia2014, takano2014, rangwa2014, viti2014,
aladro2015, martin2015, nakaji2015,
izumi2016, imanis2016, kelly2017, 
qiu2018}.

In galaxies with AGNs, high HCN/CO and HCN/HCO$^+$ intensity ratios
have been reported 
\citep[e.g.,][see also related references in 
\cite{takano2014}]{jackso1993, kohno1996, kohno2003, krips2008}.
Recently data of submillimeter lines of HCN, HCO$^+$, and CS 
became available, and consequently \cite{izumi2013, izumi2016}
concluded that the intensity ratios of HCN ($J$ = 4--3)/HCO$^+$ ($J$ = 4--3)
and/or HCN ($J$ = 4--3)/CS ($J$ = 7--6) are enhanced in circumnuclear 
disk (CND)
around AGNs,
based on data from their observations, the ALMA 
(Atacama Large Millimeter/submillimeter Array)
archive, and literatures.
In addition, recent high spatial resolution observations 
with ALMA revealed detailed distributions of 
the HCN/HCO$^+$ intensity ratio.
In Seyfert galaxies NGC 1068 and NGC 1097, 
the ratios are generally high in the CND, but the maximum 
ratios are seen at the surrounding regions of the AGN positions,
not at the AGN positions
\citep{garcia2014, viti2014, martin2015}.
\cite{salak2018} reported a similar situation in the 
starburst galaxy NGC 1808 with a weak AGN.
In Cen A the ratios are found to be 
not high ($\sim$0.5) in the central regions \citep{espada2017}. 
As shown above, the interpretation of the ratios
are still not straightforward.

Furthermore, {\it Herschel} observations opened 
new wavelength/frequency ranges of submillimeter
and THz regions.
Such observations show relatively strong intensities of reactive ions 
OH$^+$ and H$_2$O$^+$
\citep[e.g.,][]{werf2010, rangwa2011},
and 
relatively high CH/CO ratios of column density in galaxies with AGNs
\citep[][]{rangwa2014}.   

In addition, many studies of molecular abundances have been
carried out in galaxies with starbursts, such as toward NGC 253, M 82, 
\citep[e.g.,][]{mauers1991, takano1995, meier2005, martin2006, 
meier2012, aladro2013,
aladro2015, meier2015}.
Recently \cite{aladro2015} carried out systematic line survey 
observations toward eight nearby galaxies with starbursts, AGNs, and
ultra-luminous infrared emission to study the effects 
of nuclear activity.
These studies have often been carried out with unbiased line 
survey observations in the frequency axis.
The line surveys are of fundamental importance in 
astronomy not only for complete understanding of molecular 
and atomic abundances in representative sources, 
but also for finding out new observational tools 
(spectral lines) probing astrophysical phenomena.

We carried out a new line survey project
in the 3 mm wavelength region
from December 2007 to May 2012 
\citep[][]{takano2013}
as one of the legacy projects with the Nobeyama 45 m
radio telescope.
The target objects of this project include Galactic and 
extragalactic sources, and the project was subdivided into 
four sub-projects:
1. Low-mass star forming region L1527 (Yoshida  et al. accepted to PASJ),
2. Interacting shocked region L1157 B1 between the outflow and the ambient clouds
\citep{sugimu2011, yamagu2011, yamagu2012},
3. Infrared dark cloud G28.34+0.06
\citep{liu2013}, and
4. Galaxies NGC 1068, NGC 253, and
IC 342 \citep{nakaji2011, nakaji2018}.
The present sub-project surveyed lines from extragalactic objects.
The purpose of this extragalactic 
sub-project has been to study molecular abundances
in nearby galaxies with AGNs and/or starbursts.
The obtained data from the galaxies are compared one another to extract
characteristics of the effects of the AGNs and starbursts.
The additional purpose of this sub-project was to prepare
for the ALMA
early science by obtaining inventory of spectral lines
with accurate flux information:
Single-dish telescopes such as the Nobeyama 45 m can obtain
accurate flux, even if the distributions of spectral lines are
spatially more extended than the telescope beam.

Three well studied nearby galaxies, NGC 1068, NGC 253, and IC 342, were 
selected for this sub-project as 
gas-rich extragalactic sources 
\citep[e.g.,][]{young1995}
with the AGNs and/or starbursts.
The effects of the AGNs and starbursts were expected to be studied by comparing
data among the central regions in 
these three galaxies and by comparing them with chemical model 
calculations.
The three galaxies are briefly introduced below.

\begin{figure*}
 \begin{center}
\includegraphics[width=18cm]{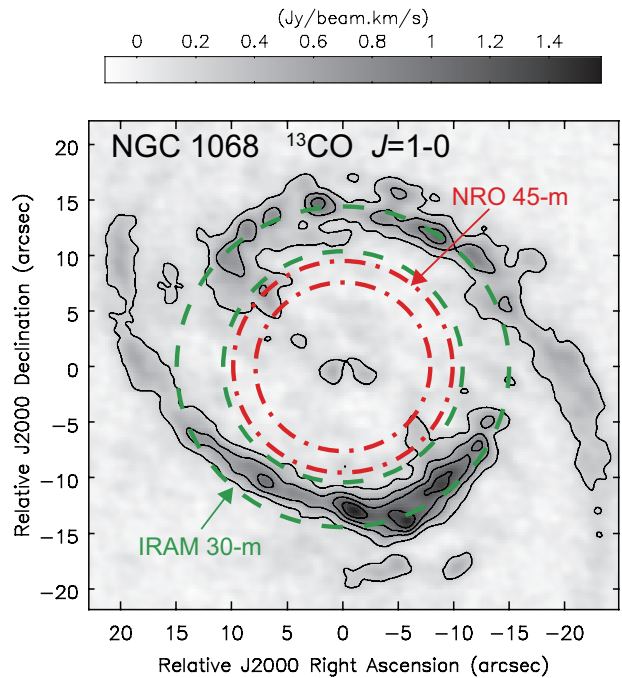}  

 \end{center}
\caption{The distribution of the 
$^{13}$CO
 $J$ = 1--0 line
  in the central region of NGC 1068 
\citep{tosaki2017}
overlaid with the beam sizes at
  the 3 mm wavelength 
  of the Nobeyama 45 m telescope (15$''$.2--19$''$.1, dot dashed lines) and 
  the IRAM 30 m telescope (21$''$--29$''$, dashed lines) \citep[][]{aladro2015}.
}\label{fig:45mbeam}
\end{figure*}

NGC 1068 (M 77) is a nearby \citep[14.4 Mpc,][]{tully1988, bland1997} 
well studied Seyfert 2 galaxy.
The CND
consists of the eastern knot
and the western knot with the separation of about 3$''$.
This CND is 
surrounded by the starburst ring/arms with a 
diameter of about 30$''$.
So far several line survey observations have been reported toward the center
of NGC 1068 
\citep[][]{
snell2011, 
costag2011, 
kamene2011, 
spinog2012, 
aladro2013,
aladro2015}
using single-dish telescopes.
In these observations emission from the CND and the  
starburst ring/arms is not well separated due to 
the relatively large telescope 
beams compared with the diameter of the starburst ring/arms
($\sim$30$''$),
except for the cases of the short wavelength regions of the
two telescopes:
 {\it Herschel} observations \citep[17$''$ 
at 194 $\mu$m,][]{spinog2012}
and the IRAM 30 m observations
\cite[22$''$ at $\sim$112 GHz,][and 21$''$ at $\sim$116 GHz, 
Aladro et al. 2013, 2015]{costag2011}.
The relatively small beam of the 45 m telescope (15$''$.2--19$''$.1) 
at the 3 mm region
can mainly observe the CND (see also the section of observations).
The beam sizes 
of the Nobeyama 45 m and the IRAM 30 m radio telescopes
are overlaid on the $^{13}$CO image of NGC 1068 in 
figure \ref{fig:45mbeam}  
for comparison. 

NGC 253 is a nearby \citep[3.5 Mpc, e.g.,][]{rekola2005, mouhci2005} 
almost edge-on barred spiral galaxy with 
prototypical starbursts.
It has an exceptionally rich gas with many molecular species in high abundance
\citep[e.g.,][]{mauers1991}.
The first extragalactic line survey was reported by \cite{martin2006}
toward the center of this galaxy at the 2 mm wavelength 
region with the IRAM 30 m radio telescope.
Then, several line survey observations have been reported toward the center
of NGC 253 
using single-dish telescopes and ALMA
\citep[][]{
snell2011, 
rosenb2014,
meier2015,
aladro2015}.
Although \cite{snell2011} 
and \cite{aladro2015} 
observed with the single-dish telescopes,
the FCRAO 14 m and the IRAM 30 m, respectively,
at the same wavelength of 3 mm with our observations, 
our beam (15$''$.2--19$''$.1) is smaller than their beams
(47$''$--70$''$ and 21$''$--29$''$ for 
the FCRAO 14 m and the IRAM 30 m, respectively).
 
IC 342 is a nearby \citep[3.93 Mpc,][]{tikhon2010} 
almost face-on barred spiral galaxy with 
prototypical starbursts.
This galaxy is also known to have rich molecular gas
\citep[e.g.,][]{henkel1988}. 
Limited line survey observations have been reported toward the center
of IC 342 
\citep[][]{
snell2011, 
rigopo2013}
using single-dish telescopes.
Since IC 342 is situated in the northern sky, it cannot be observed with ALMA,
but it can be observed, for example, with NOEMA
(Northern Extended Millimeter Array).
Although \cite{snell2011}
observed  with the FCRAO 14 m radio telescope 
at the same wavelength of 3 mm with our observations, 
our beam is again smaller ($\sim$1/3) than their beams, and 
our velocity resolution is about 10 times higher than their 
resolution ($\sim$100 km s$^{-1}$).
\cite{rigopo2013} observed with  {\it Herschel} at the wavelength 
of 196--671 $\mu$m.
The properties of NGC 1068, NGC 253, and IC 342 are 
listed  in table 1.

The initial results of NGC 1068 
and this entire line survey project
were already reported by  
\cite{nakaji2011} 
and
\cite{takano2013}, respectively.
In this article, we present obtained data for the three galaxies
with the 45 m telescope. 
Then, we discuss results immediately recognized from the data.
The analyses employing rotational diagrams and discussion on the 
molecular abundances were already reported in a separate paper by \cite{nakaji2018} 
(hereafter ``analysis paper'').   

\section{Observations}

The observations were carried out with the 45 m telescope at
Nobeyama Radio Observatory 
(NRO)\footnote{Nobeyama Radio Observatory is a branch of the National 
Astronomical Observatory of Japan, National Institutes of Natural Sciences.}
between  February 2009 and May 
2011 (three observational seasons).
The total allocated time was $\sim$500 hrs.
Among them $\sim$204 hrs were used for the main observations for
the galaxies 
(about 87, 41, and 76 hrs for NGC 1068, NGC 253, and 
IC 342, respectively)
excluding the time for receiver tunings, pointings of the telescope, 
calibration, recovery from troubles of the system, and bad weather.
The frequency covered was from $\sim$85 to $\sim$116 GHz:
There are some gaps in the frequency coverage due to reasons of
frequency settings.
We tried to place the gaps in frequency regions without
expected significant lines (except for HN$^{13}$C, see
the section \ref{impnon}).
The dual-polarization and sideband-separating (2SB) receivers T100
for the 3 mm region \citep{nakaji2008} were used:
They can observe both
linear polarizations (T100H and T100V) with two
sidebands simultaneously, and with 
higher sensitivity than the previous observations 
with the old SIS receivers.
The system temperature was typically 150--300 K including the
atmospheric noise depending on elevation, weather, and 
frequency.
The image-sideband rejection ratio was typically $>$10 dB.
The ratio was measured by injecting artificial signal from the 
top of the receiver optics after each tuning of the
receivers \citep{nakaji2010}.
The beam sizes were 15$''$.2--19$''$.1
 (half power beam width: HPBW)
 for 115--86 GHz.
The intensity calibration to obtain antenna temperature ($T_{\rm A}^*$)
was carried out with the chopper-wheel method.
We employed the main beam efficiencies of the 
T100 receivers for each year to convert the intensity
scale to main beam temperature ($T_{\rm mb}$).
The variation of the efficiency during our 
observational period was not significant: The 
main beam
efficiencies
in the final year of our observations
at 86, 110, and 115 GHz were 42, 42, and 36\%, respectively,
for T100H, and 43, 42, and 
36\%, respectively, for T100V.
 
Before December 2010 the backend used was digital spectrometers 
AC45 \citep{sorai2000}.
Eight spectrometers with an instantaneous
bandwidth of 512 MHz each
and with a resolution of 605 kHz were used simultaneously.
From December 2010
a new IF system, new AD  converters 
(4 GHz sampling rate with 3 bit), new digital spectrometers SAM45
(sixteen spectrometers with an instantaneous 
bandwidth of 
$\sim$1.6 GHz each at the maximum bandwidth), 
and new related softwares were available.  
This new system \citep{kuno2011, iono2012}
accelerated our survey.  
SAM45 was made based on the technology of the correlator for the 
ALMA Atacama Compact Array (Morita Array) \citep{kamaza2012}. 
The resolution of SAM45 was set to be 488.28 kHz.
In addition,
the Doppler tracking of both of the sidebands for the 2SB type
receivers 
was
carried out by software after the data acquisition.
Such software was implemented when the 2SB type receivers
were installed \citep{takaha2010}.

The central position of each galaxy was observed.
The coordinates and the systemic velocities employed were 
RA(J2000.0) = \timeform{2h42m40s.798}\footnote{In \cite{schinn2000}
this value
was referred from \cite{muxlow1996}, where  
a slightly different value of 
\timeform{2h42m40s.7098} was reported.}, 
Dec(J2000.0) = \timeform{-00D00'47".938},
and 1150 km s$^{-1}$ for NGC 1068 \citep{schinn2000},
RA(J2000.0) = \timeform{00h47m33s.3}, 
Dec(J2000.0) = \timeform{-25D17'23"} \citep{martin2006},
and 230 km s$^{-1}$ for NGC 253,
RA(J2000.0) = \timeform{03h46m48s.9}, 
Dec(J2000.0) = \timeform{68D5'46.0"} \citep{falco1999},
and 32 km s$^{-1}$ \citep{crosth2000} for IC 342.
These parameters are summarized in table 1.
The position switching was employed.
The integration time was 10--20 s both for ON and OFF positions.
The OFF positions were +5$'$ of azimuthal angle
for the three galaxies.
The telescope pointing was checked every 1--1.5 hours using 
the nearby SiO maser sources ($v$ = 1 and/or 2, $J$ = 1--0):
$o$-Cet for NGC 1068, R-Aqr for NGC 253, and 
T-Cep and IRC+60092 for IC 342.
The pointing deviations were typically within 5$''$.

\section{Data Reduction}

The data reduction was carried out with the 
software package for the spectral lines NewStar
\citep{ikeda2001}.
All individual scans were visually 
inspected, and
the bad data (e.g., bad baselines) were flagged
manually. 
Then the
data were integrated, baseline subtracted, and binned
to obtain the final spectra. 
Linear baselines were usually subtracted.
The lines were Gaussian fitted to obtain intensity,
line of sight velocity, and width (full width at half intensity).
The integrated intensities were obtained by numerically
summing intensity at each spectral channel with 
significant intensity above the baselines.

The obtained spectra are presented in this article.
The spectra were already
analyzed to obtain rotational
temperatures and column densities in the analysis paper, where the beam
dilutions were taken into account.

\section{Results}

\begin{figure*}
 \begin{center}
   \includegraphics[width=16cm]{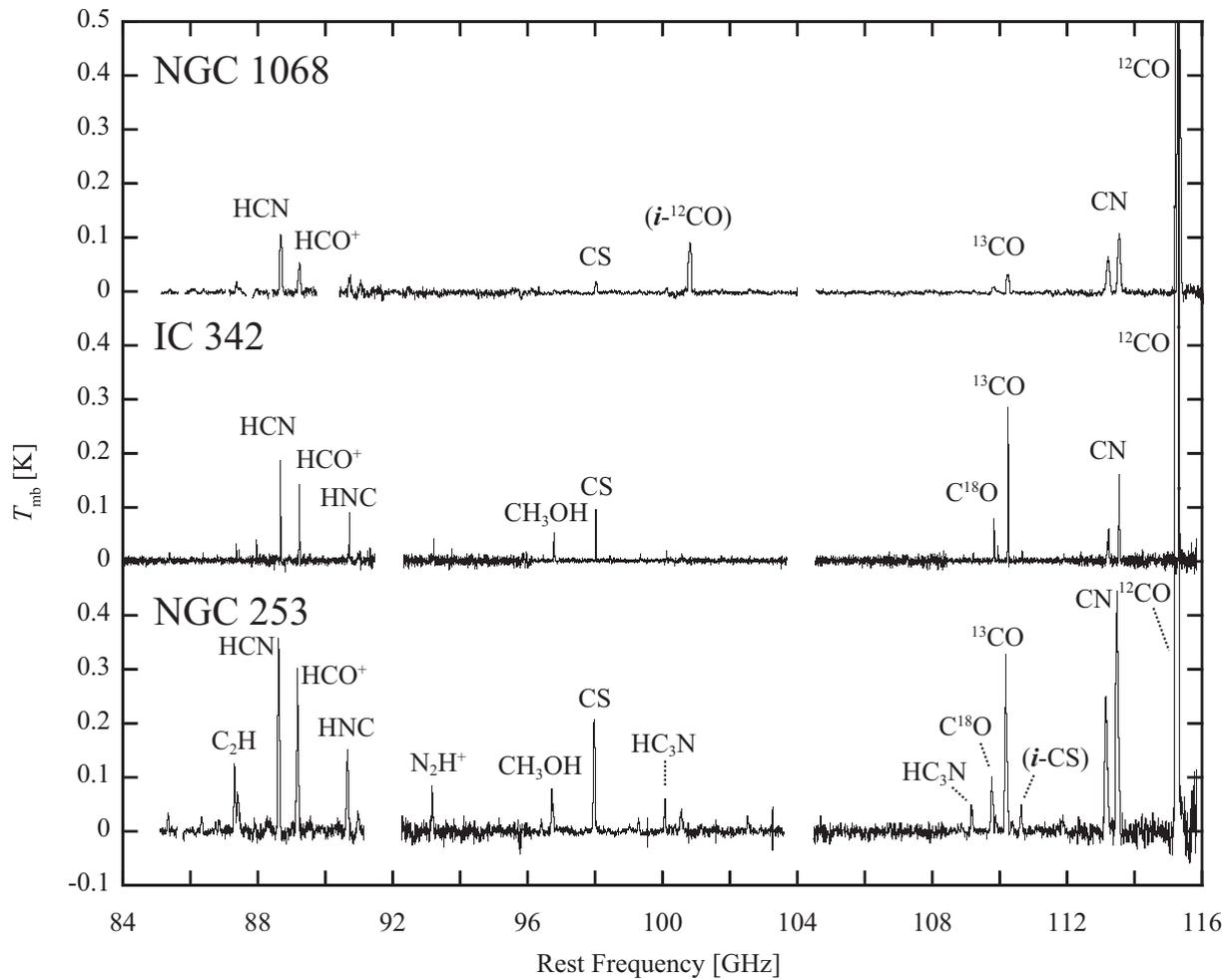} 
 \end{center}
\caption{The compressed spectra obtained from the line survey 
  observations toward NGC 1068, NGC 253, and IC 342.
HCN, HCO$^+$, and CN line intensities relative to $^{13}$CO are strong
in NGC 1068 among the three galaxies.
A line with {\it 'i'} and a name of the molecule in parentheses indicates 
a signal of the corresponding molecule leaked
from the other sideband of the receiver.
}\label{fig:3spectra}
\end{figure*}

Spectra from $\sim$85 GHz to $\sim$116 GHz
toward the nearby external
galaxies NGC 1068, NGC 253, and IC 342 were obtained.
The whole compressed spectra of the three galaxies are presented in
figure \ref{fig:3spectra}.
The spectra were made with 
velocity resolutions of 
20 km s$^{-1}$ for NGC 1068 and NGC 253, and
10 km s$^{-1}$ for IC 342.
As a result, the achieved rms noise levels (in $T{\rm_A^*}$) are 
1.2--2.6 mK for NGC 1068,
1.8--4.8 mK for NGC 253, and
0.8--2.5 mK for IC 342.

\begin{figure*}
 \begin{center}
  \includegraphics[width=16cm]{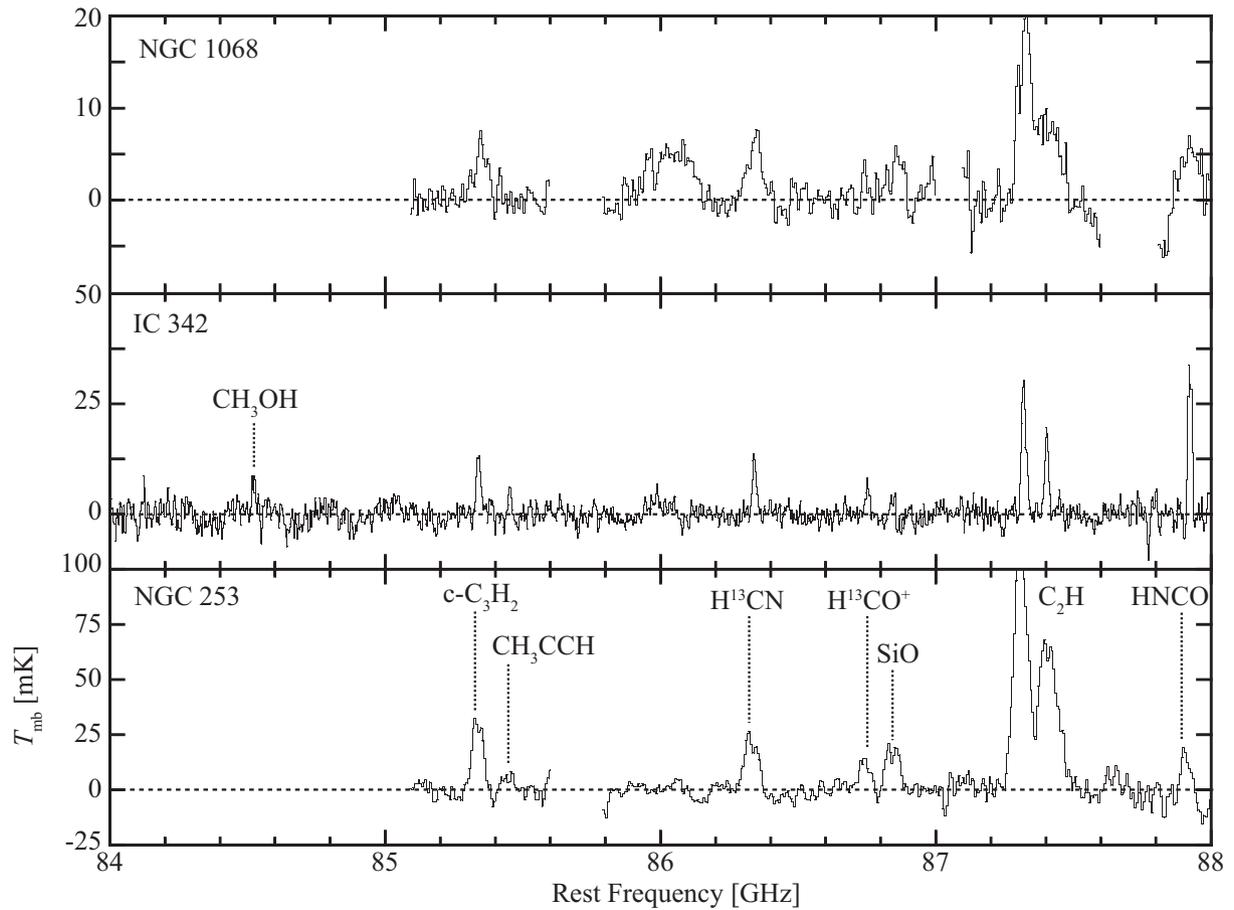} 
 \end{center}
\caption{The spectra obtained from the line survey 
  observations toward NGC 1068, NGC 253, and IC 342 at 84-88 GHz
are presented.
A broad feature at $\sim$86.0 GHz in the spectra of NGC 1068 is caused by
baseline fluctuation.
}\label{fig:all1}
\end{figure*}

\begin{figure*}
 \begin{center}
\includegraphics[width=16cm]{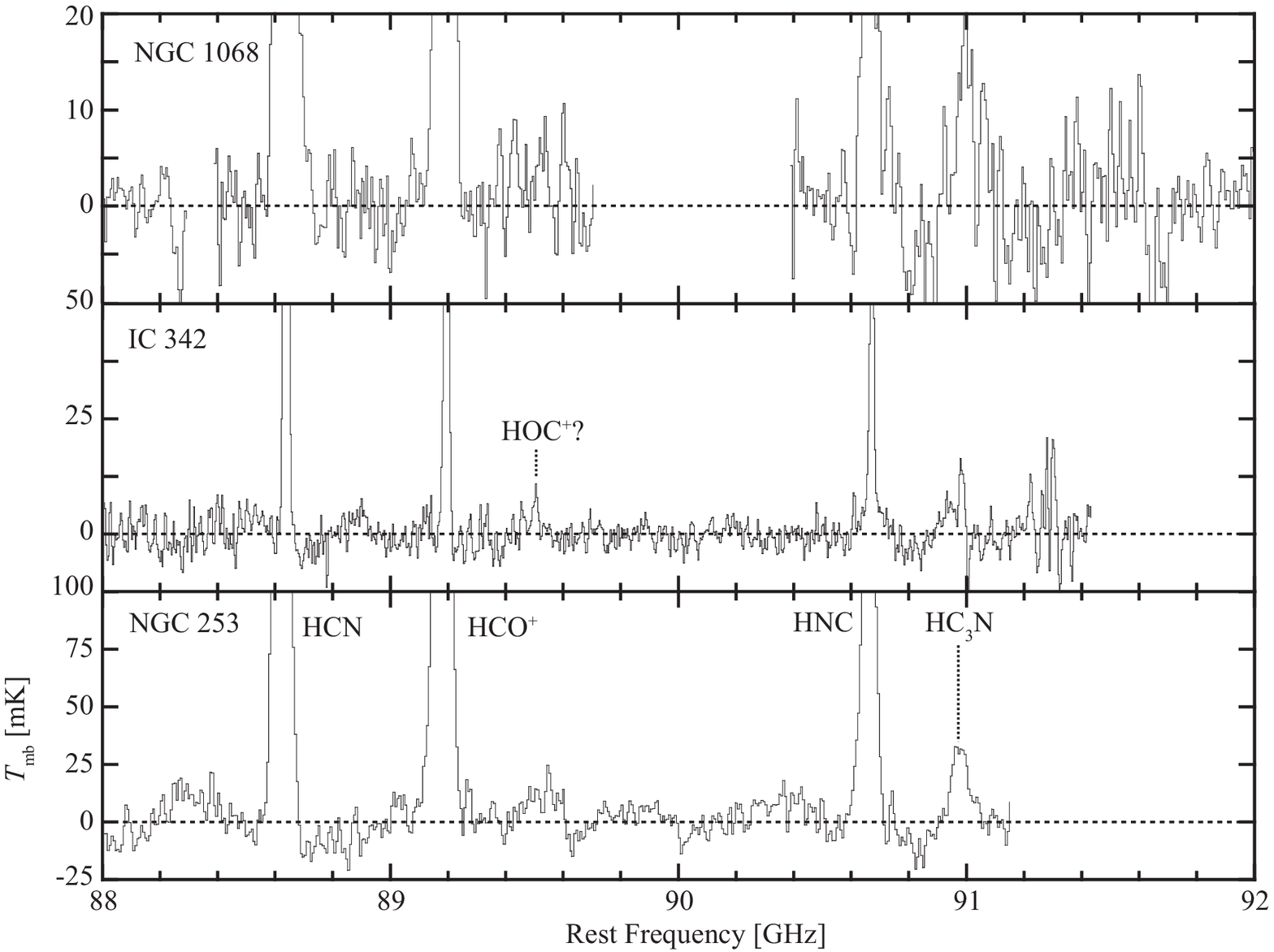}
 \end{center}
\caption{The spectra obtained from the line survey 
  observations toward NGC 1068, NGC 253, and IC 342: 88-92 GHz.}\label{fig:all2}
\end{figure*}

\begin{figure*}
 \begin{center}
   \includegraphics[width=16cm]{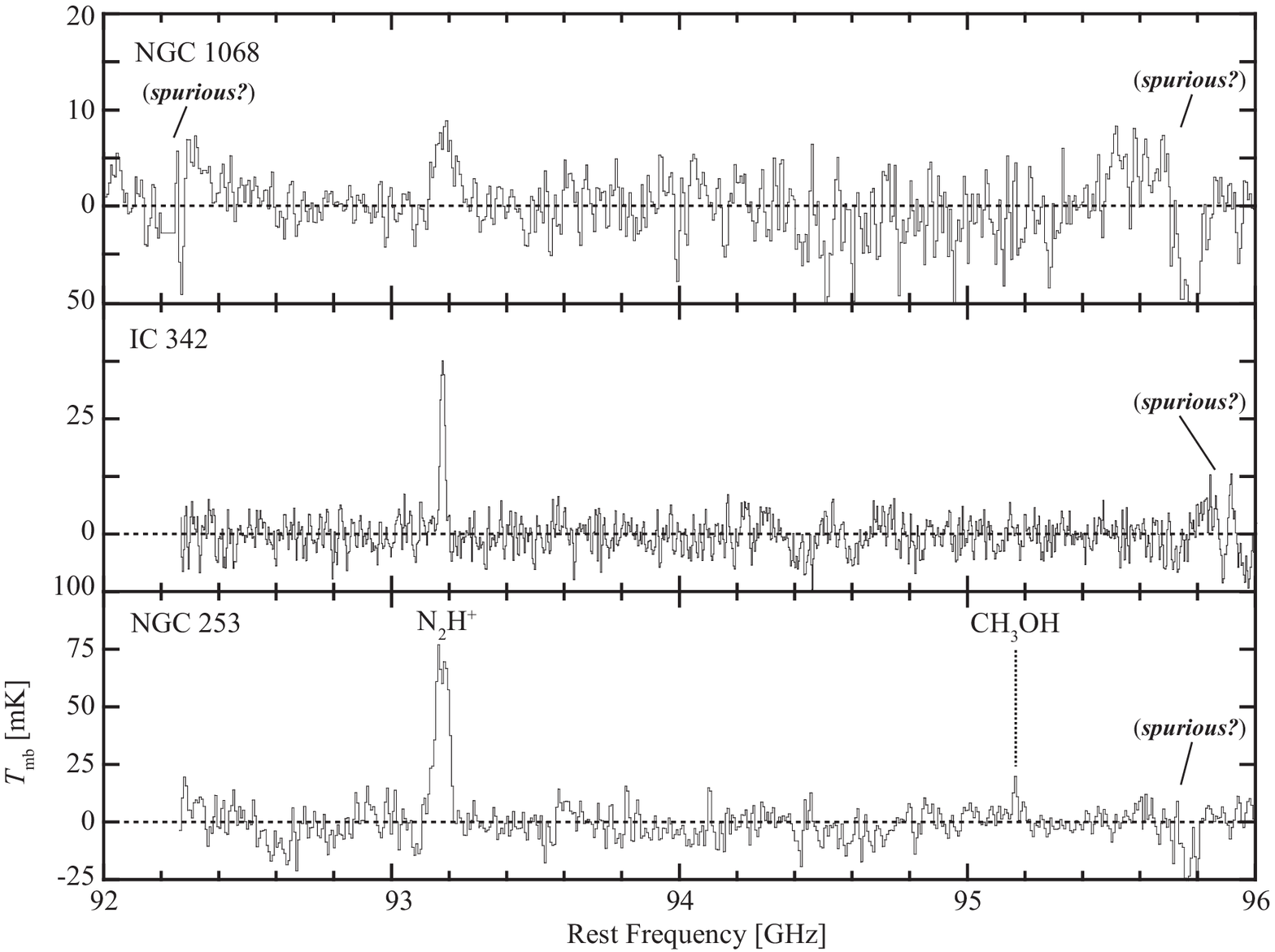}
 \end{center}
\caption{The spectra obtained from the line survey 
  observations toward NGC 1068, NGC 253, and IC 342: 92--96 GHz.}\label{fig:all3}
\end{figure*}

\begin{figure*}
 \begin{center}
     \includegraphics[width=16cm]{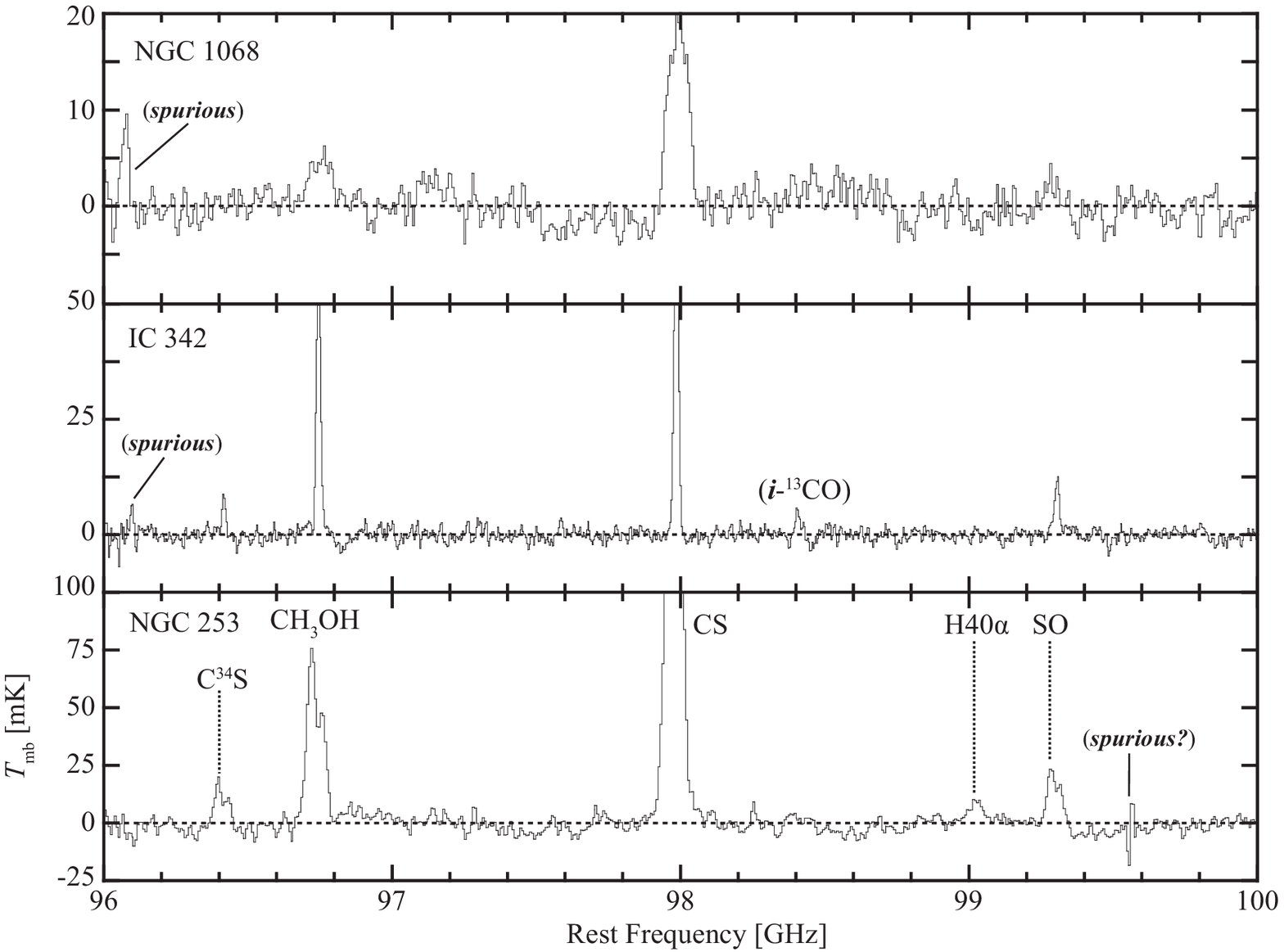} 
 \end{center}
\caption{The spectra obtained from the line survey 
  observations toward NGC 1068, NGC 253, and IC 342: 96--100 GHz.
A line with {\it 'i'} and a name of the molecule in parentheses indicates 
a signal of the corresponding molecule leaked
from the other sideband of the receiver.
}\label{fig:all4}
\end{figure*}

\begin{figure*}
 \begin{center}
  \includegraphics[width=16cm]{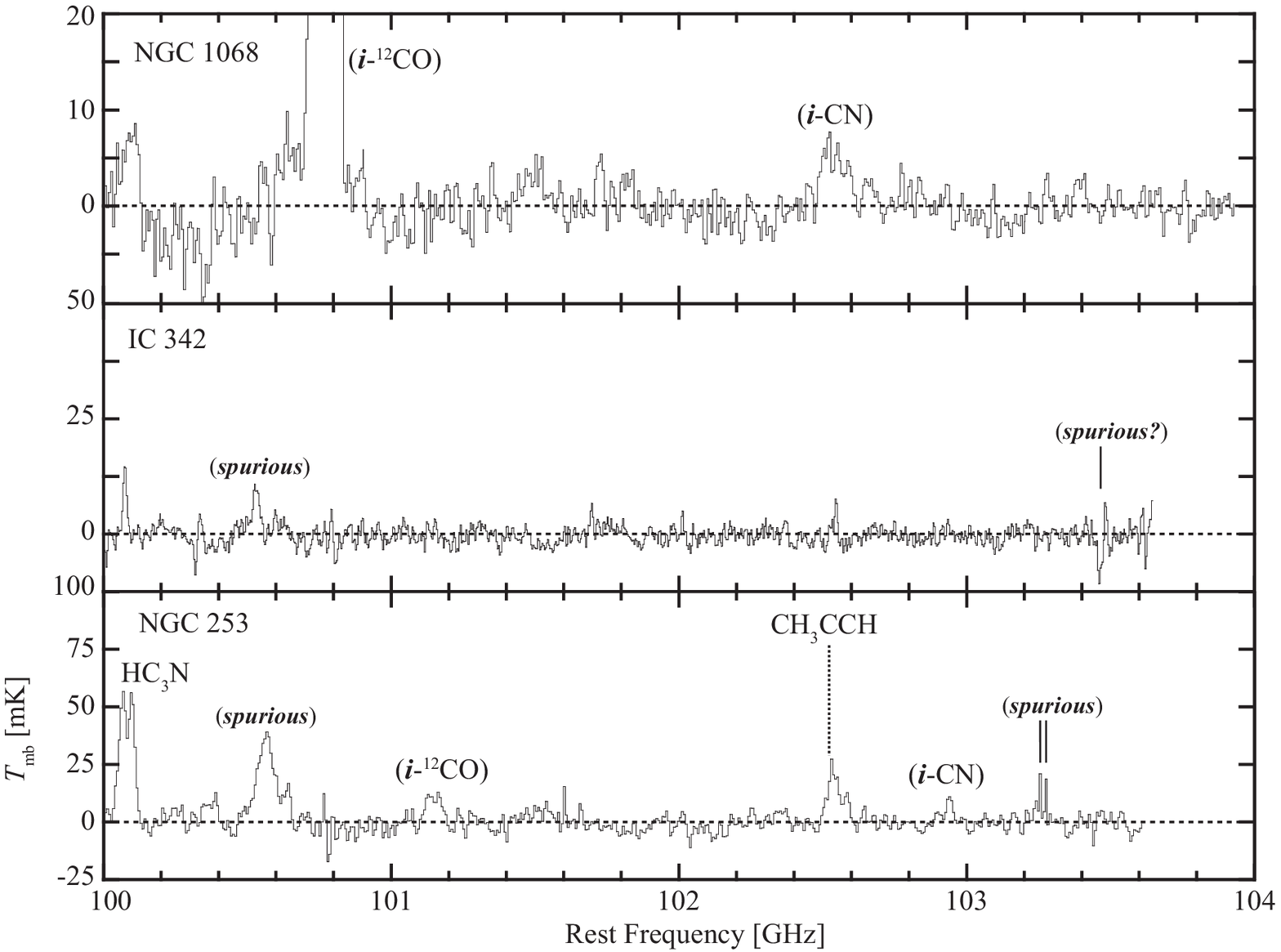} 
 \end{center}
\caption{The spectra obtained from the line survey 
  observations toward NGC 1068, NGC 253, and IC 342: 100--104 GHz.
A line with {\it 'i'} and a name of the molecule in parentheses indicates 
a signal of the corresponding molecule leaked
from the other sideband of the receiver.
}\label{fig:all5}
\end{figure*}

\begin{figure*}
 \begin{center}
  \includegraphics[width=16cm]{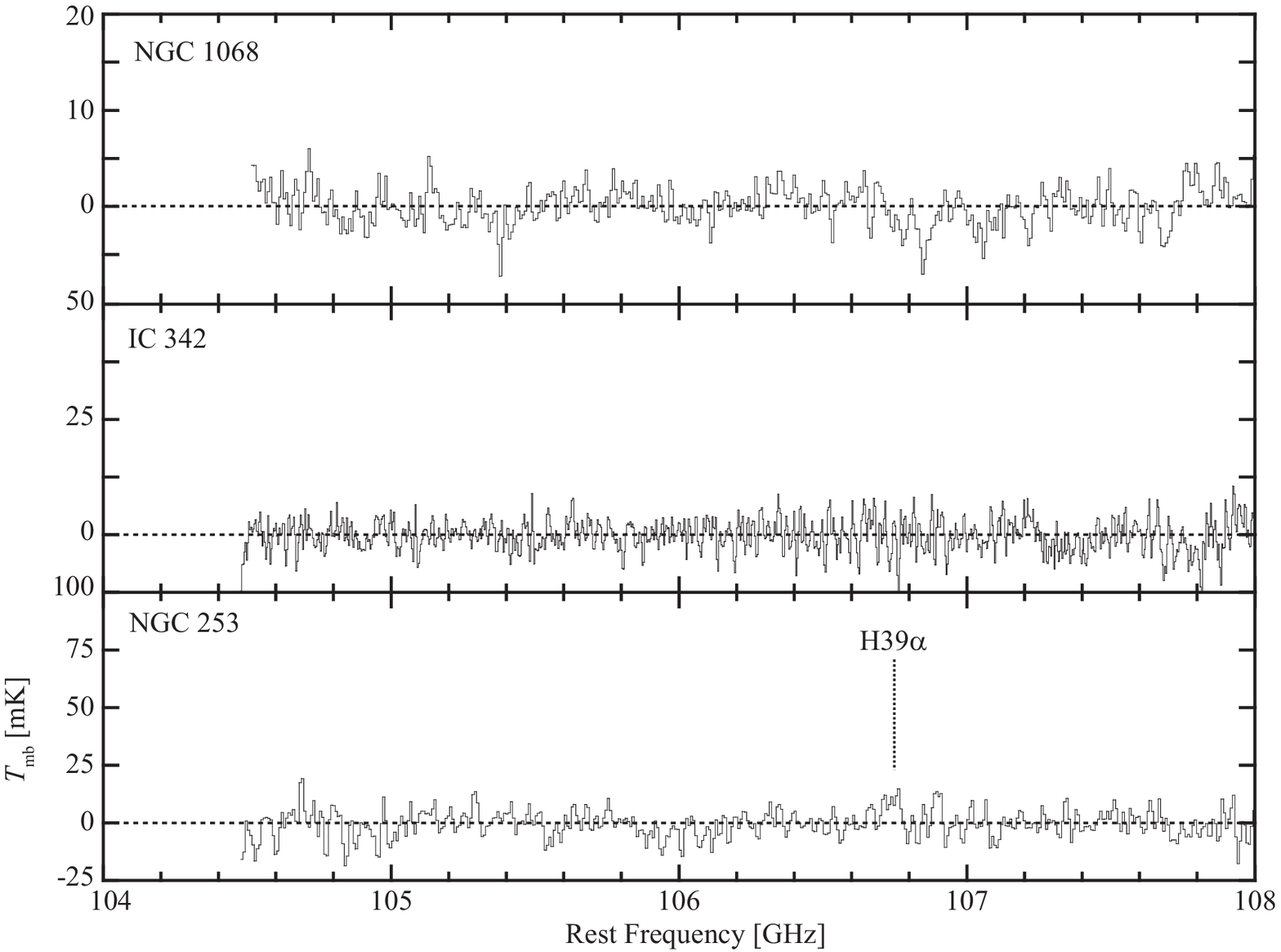} 
 \end{center}
\caption{The spectra obtained from the line survey 
  observations toward NGC 1068, NGC 253, and IC 342: 104-108 GHz.}\label{fig:all6}
\end{figure*}

\begin{figure*}
 \begin{center}
  \includegraphics[width=16cm]{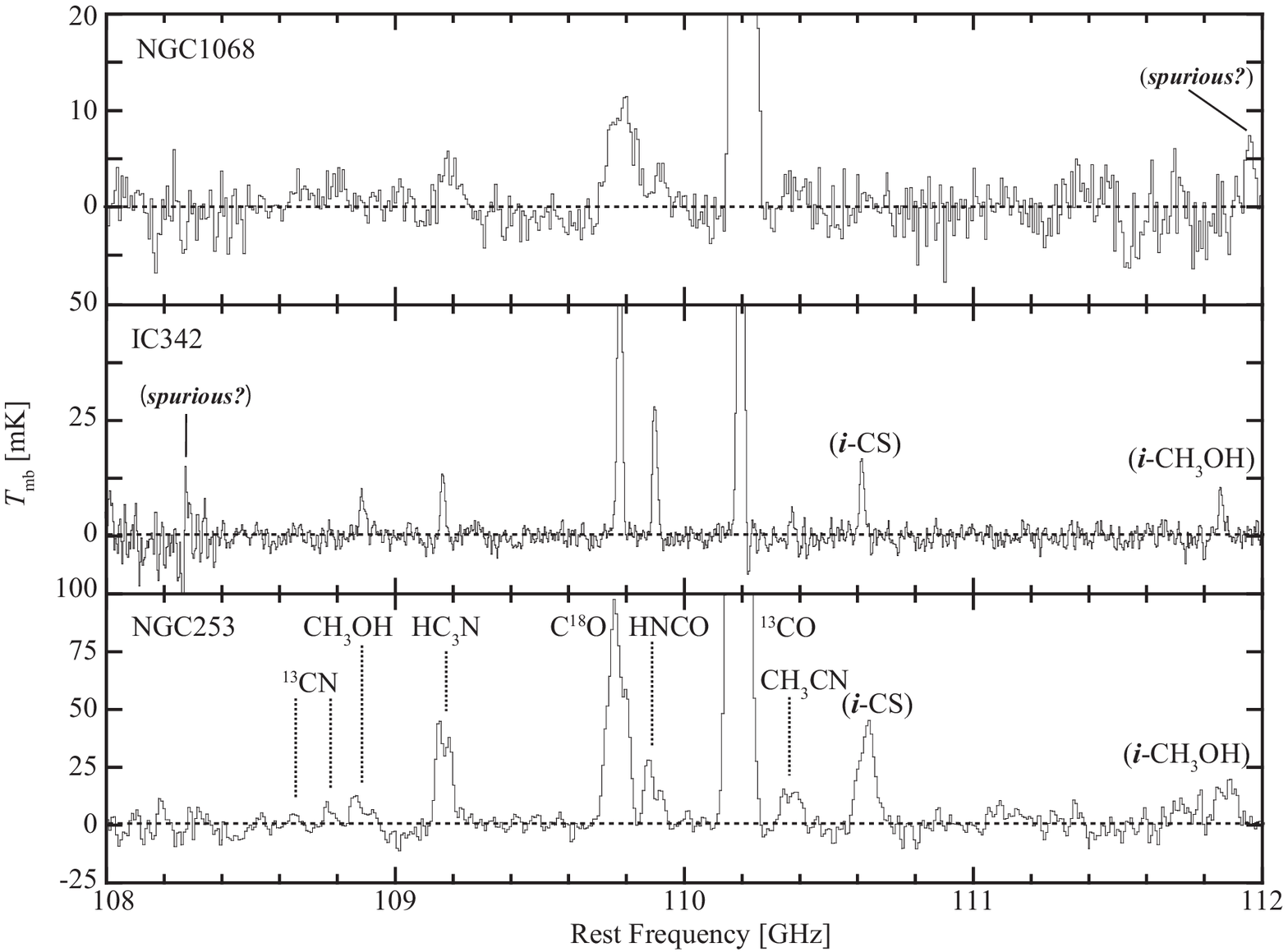} 
 \end{center}
\caption{The spectra obtained from the line survey 
  observations toward NGC 1068, NGC 253, and IC 342: 108--112 GHz.
A line with {\it 'i'} and a name of the molecule in parentheses indicates 
a signal of the corresponding molecule leaked
from the other sideband of the receiver.
}\label{fig:all7}
\end{figure*}

The spectra with 4 GHz width are presented in figures \ref{fig:all1} to 
 \ref{fig:all8}.
The line parameters such
as intensity are listed in tables \ref{tab:param1} to \ref{tab:param3}.
The rest frequencies were obtained from 
\cite{lovas2004}.
During the line identification
(see the section \ref{line-i}), we found lines of
$^{12}$CO, $^{13}$CO, CN, CS, and CH$_3$OH
leaked from the other sideband of the receivers.
They are indicated in the spectra.
In addition the AD converters at the period of
our observations 
generated spurious lines (a few moving features during observations
along the 
frequency axis per 
each 1.6 GHz instantaneous bandwidth).
Significant features of such spurious lines 
(sometimes broad features after integration)
are also indicated in the spectra.

\subsection{Line identification\label{line-i}}
\subsubsection{Detected Lines\label{identi}}
The detected 
lines were identified based on the literatures of spectral line
 surveys referred in the 
introduction, and also on databases of molecular spectroscopy,
CDMS, 
JPL Catalog \citep{picket1998}, 
NIST Recommended Rest Frequencies \citep[e.g.,][]{lovas2004}, 
and Splatalogue
\citep[e.g.,][]{remija2016}.
We judge detections based not only on the intensity, 
but also on the quality of the data such as 
characteristics of the noise (random or not) and
characteristics of the baseline (flat or fluctuated).

The numbers of lines
detected are 25, 34, and 31 for
NGC 1068, NGC 253, and IC 342, respectively.
As a result, the numbers of atomic and molecular species 
(distinguishing isotopologues) identified
are 19, 24, and 22 for
NGC 1068, NGC 253, and IC 342, respectively.
Atomic hydrogen is the only detected atomic species
as recombination lines, and it is detected only in NGC 253. 

The detected molecules in NGC 1068 for the first time in this survey
(C$_2$H, cyclic-C$_3$H$_2$, and H$^{13}$CN)
were already reported in our
paper of the initial results \citep{nakaji2011}.
In NGC 253 molecules detected in this survey have already 
been reported in literatures
\citep[e.g.,][]{mauers1991, aladro2015}.
In IC 342 cyclic-C$_3$H$_2$, SO, and C$^{17}$O are 
detected for the first time.

\subsubsection{Tentative detections\label{tentat}}
\begin{figure*}
 \begin{center}
     \includegraphics[width=16cm]{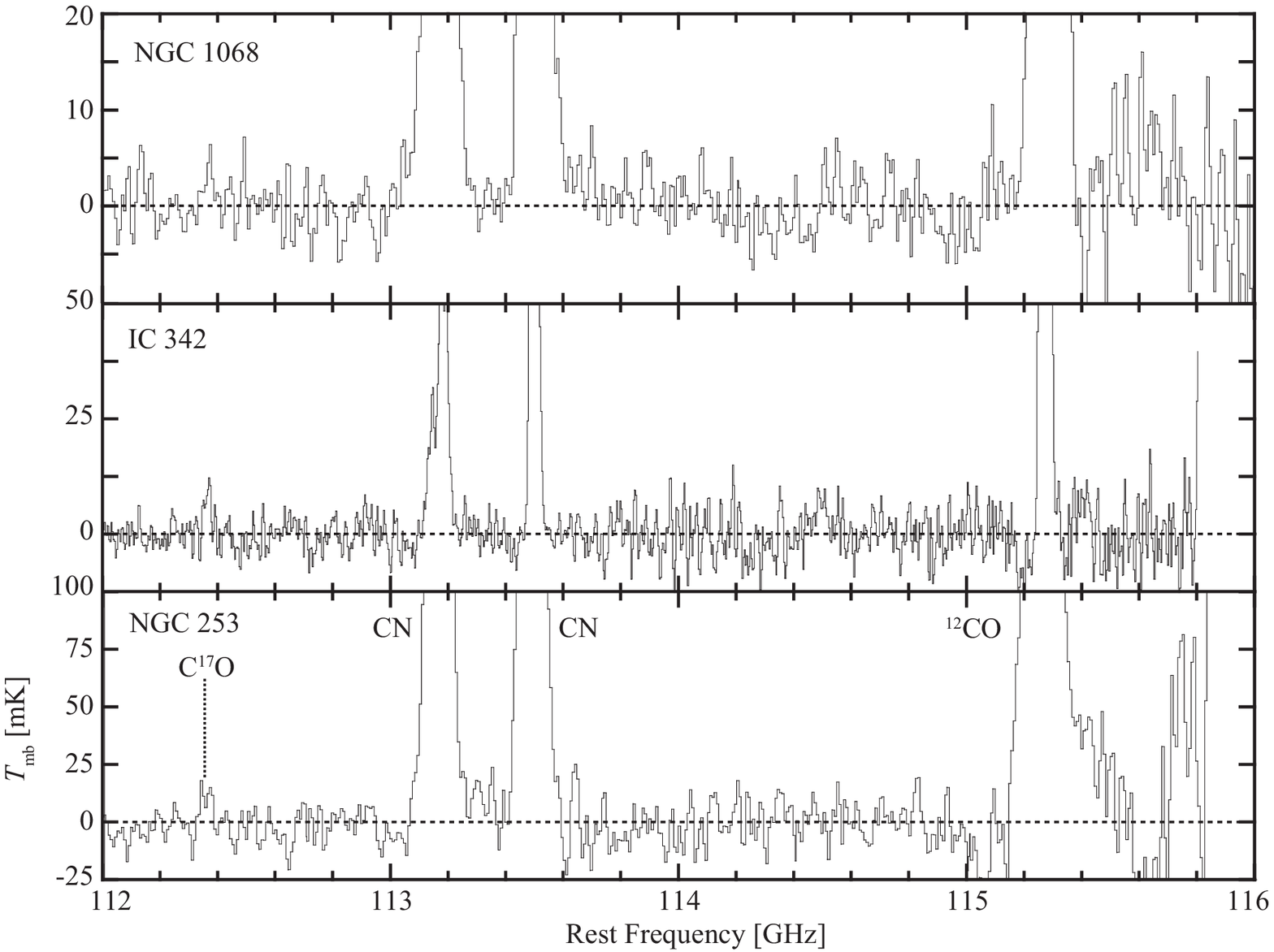}  
 \end{center}
\caption{The spectra obtained from the line survey 
  observations toward NGC 1068, NGC 253, and IC 342: 112--116 GHz.}\label{fig:all8}
\end{figure*}

In NGC 1068 a 
possible weak 
and broad
emission feature is seen at the frequency of 
the $J$ = 1--0 transition of C$^{17}$O (112358.988 MHz)
(figure \ref{fig:all8}).
Since the signal-to-noise ratio (SN) 
is low ($\sim$2), this is a tentative detection.
In NGC 253 the spectral profile of
the $J$ = 1--0 transition of $^{12}$CO has a shoulder
at the higher frequency side
(figure \ref{fig:all8}).
It may correspond to the $J$ = 5/2--3/2 transition
($^2\Pi_{1/2}$) of NS
(115556.253 MHz),
but the characteristics of the baseline seems not good.
It is, therefore, a tentative detection in our spectra.
Actually this line is detected with ALMA in NGC 253
\citep{meier2015}.
In IC 342 a weak emission feature is seen at the frequency of 
the $J$ = 1--0 transition of HOC$^+$
(89487.414 MHz)
(figure \ref{fig:all2}).
Since the SN 
is low ($\sim$2), this is a tentative detection.
It is not clear whether this line is detected 
in NGC 1068 and NGC 253 in our spectra due to the large noise
and bad baseline, respectively.
These lines are not counted as the identified line in the section \ref{identi}

\subsubsection{Important non-detections\label{impnon}}
CH$_3$CCH is known to be abundant 
in some starburst
galaxies (see the section \ref{dis-imp}).
In this work it is not detected in NGC 1068,
but it is detected in NGC 253 and IC 342.
The non-detection in NGC 1068 
is already reported by \cite{aladro2013, aladro2015}
with the IRAM 30 m telescope.
In our observations with the NRO 45 m telescope, this result 
was
further 
studied
toward the CND with the smaller beam.
Very recently \cite{qiu2018} reported
detection of CH$_3$CCH  ($J_K$ = 5$_0$-4$_0$)
with the IRAM 30 m telescope
by conducting deep observations.
These results are discussed in the section
\ref{dis-imp}.

H$^{13}$CO$^+$ is an interesting species 
to compare its intensity 
to that of H$^{13}$CN to estimate the HCN/HCO$^+$
intensity ratio to study the power source of 
galaxies as mentioned in the introduction.
It is not detected in NGC 1068,
but it is detected in NGC 253 and IC 342.
Its detection in NGC 1068 is reported by \cite{aladro2013},
though the line is partially blended with those of SiO ($J$ = 2--1) and 
possibly HCO (1$_{1, 0}$--0$_{1, 0}$).
In our data the SiO ($J$ = 2--1) line is detected, but the feature
of H$^{13}$CO$^+$ is not clear.
This result causes high H$^{13}$CN/H$^{13}$CO$^+$ intensity
ratio (see the section \ref{r-hcnhcop}).
The upper limits of the integrated intensities
($I$(upper limit)) of 
CH$_3$CCH  ($J_K$ = 5$_K$-4$_K$) and 
H$^{13}$CO$^+$  ($J$ = 1--0) in NGC 1068
are listed in table \ref{tab:param1}.
However, the upper limit of the $J_K$ = 6$_K$-5$_K$ transition of CH$_3$CCH
is not listed, because this line is 
contaminated by an image line from the other sideband
of the receiver,
which is caused by the finite image-band suppression 
of the T100 receiver.
The upper limits were calculated with the following formula
\citep[cf.][]{aladro2015}
\begin{equation}
I({\rm upper\: limit}) = {\rm rms} \times \sqrt{{\rm FWZI} \times \Delta v},
\end{equation}
where 
rms is root-mean-square of the noise close to the line frequency, 
FWZI is Full Width at Zero Intensity (estimated to be 400 km s$^{-1}$),
and   $\Delta v$ is the velocity resolution of the spectra
(20 km s$^{-1}$).

In NGC 1068 \cite{aladro2013} 
detected the $J$ = 5/2--3/2 transition
($^2\Pi_{1/2}$) of NS with the IRAM 30 m telescope.
In our spectra the frequency region of this line is noisy, 
and the line is not seen.
In addition,  \cite{aladro2015} tentatively detected HCO (1$_{1, 0}$--0$_{1, 0}$),
HN$^{13}$C ($J$ = 1--0), HOC$^+$ ($J$ = 1--0),
and C$^{34}$S ($J$ = 2--1)  in NGC 1068
with the IRAM 30 m telescope.
The lines of HCO are not seen in our spectra, though they are
generally not easy to detect due to the blending with other lines
as mentioned above.
In the case of HN$^{13}$C the frequency of this line 
(87090.859 MHz)
is in the gap of our spectra. 
HOC$^+$ is not detected in our spectra as mentioned above
in the section \ref{tentat}.
C$^{34}$S is not detected either in our spectra, though the spectra
show relatively low noise and stable baseline.
The emission of CS ($J$ = 2--1) is known to be distributed both 
in the CND and the starburst ring based on interferometric observations
(see the section \ref{d-int-ratio}). 
Therefore, it is probable that the IRAM 30 m telescope
is advantageous for the detection of C$^{34}$S ($J$ = 2--1), 
because its beam covers the emission 
both from the CND and the 
starburst ring.

\subsection{Characteristics of integrated intensity (Normalized by
CS or $^{13}$CO)}
Here we compare the integrated intensities among the three galaxies 
to see their immediately recognized characteristics.
For this comparison, the integrated intensities normalized by the 
integrated intensity of CS ($J$ = 2--1) or 
$^{13}$CO ($J$ = 1--0) were used to cancel the 
difference in the amount of gas in these galaxies.
CS is one of the typical molecules to trace dense molecular gas.
On the other hand,  $^{13}$CO ($J$ = 1--0) traces gas including relatively
low density gas \citep[e.g.,][]{wilson2013}, and its optical depth is 
much lower than that of
 $^{12}$CO.

\begin{figure*}
 \begin{center}
\includegraphics[width=16cm]{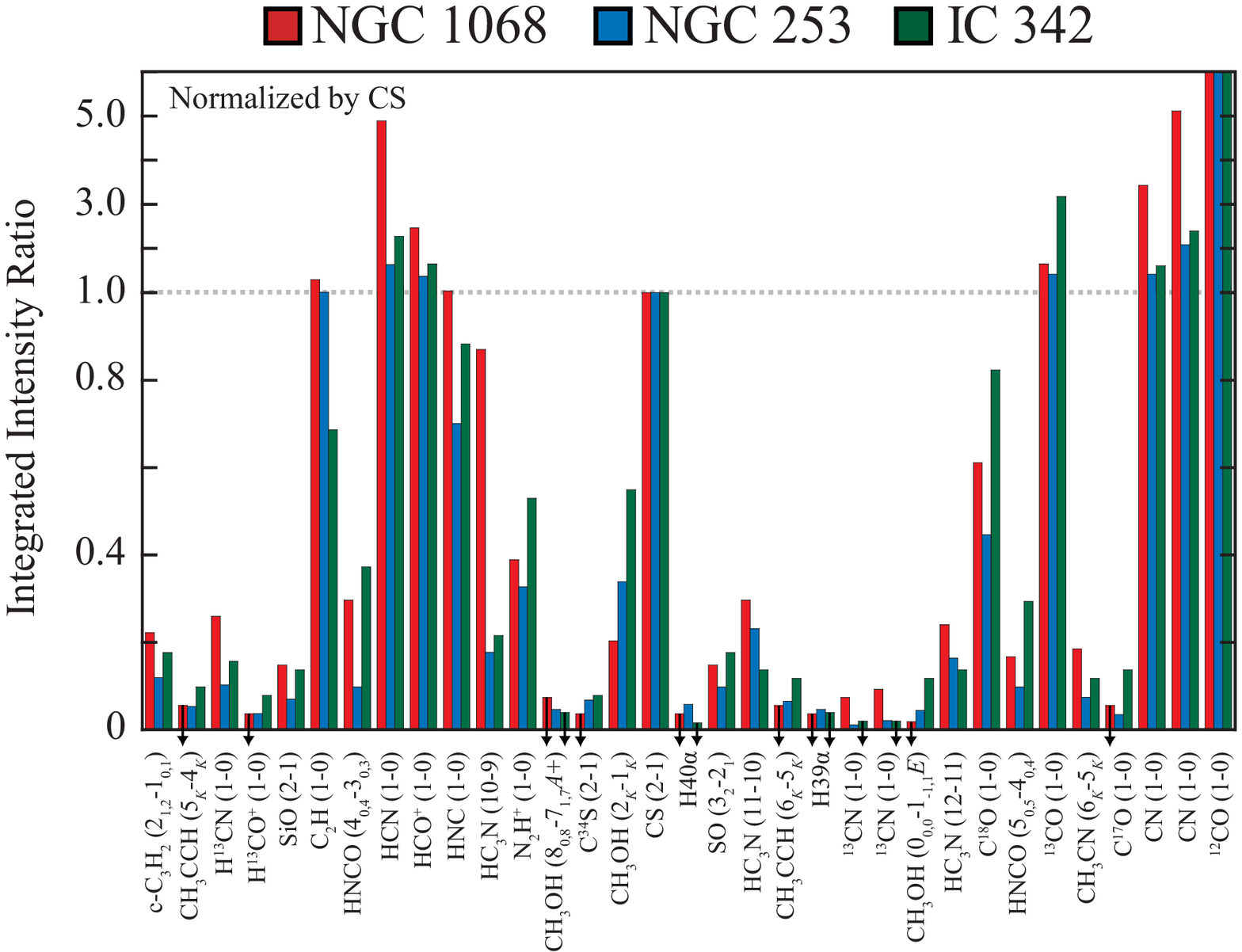} 
 \end{center}
\caption{Comparison of integrated intensity toward NGC 1068, NGC 253, and IC 342
(normalized by CS): The horizontal axis shows names of molecules and their
transitions
in the order of their transition frequencies. 
The vertical axis shows integrated intensity ratios in two
linear scales (0 to 1.0 below the dotted line, and 1.0 to 6.0 above
the dotted line).
The arrows indicate upper limits.}\label{fig:intint1}
\end{figure*}

\begin{figure*}
 \begin{center}
\includegraphics[width=16cm]{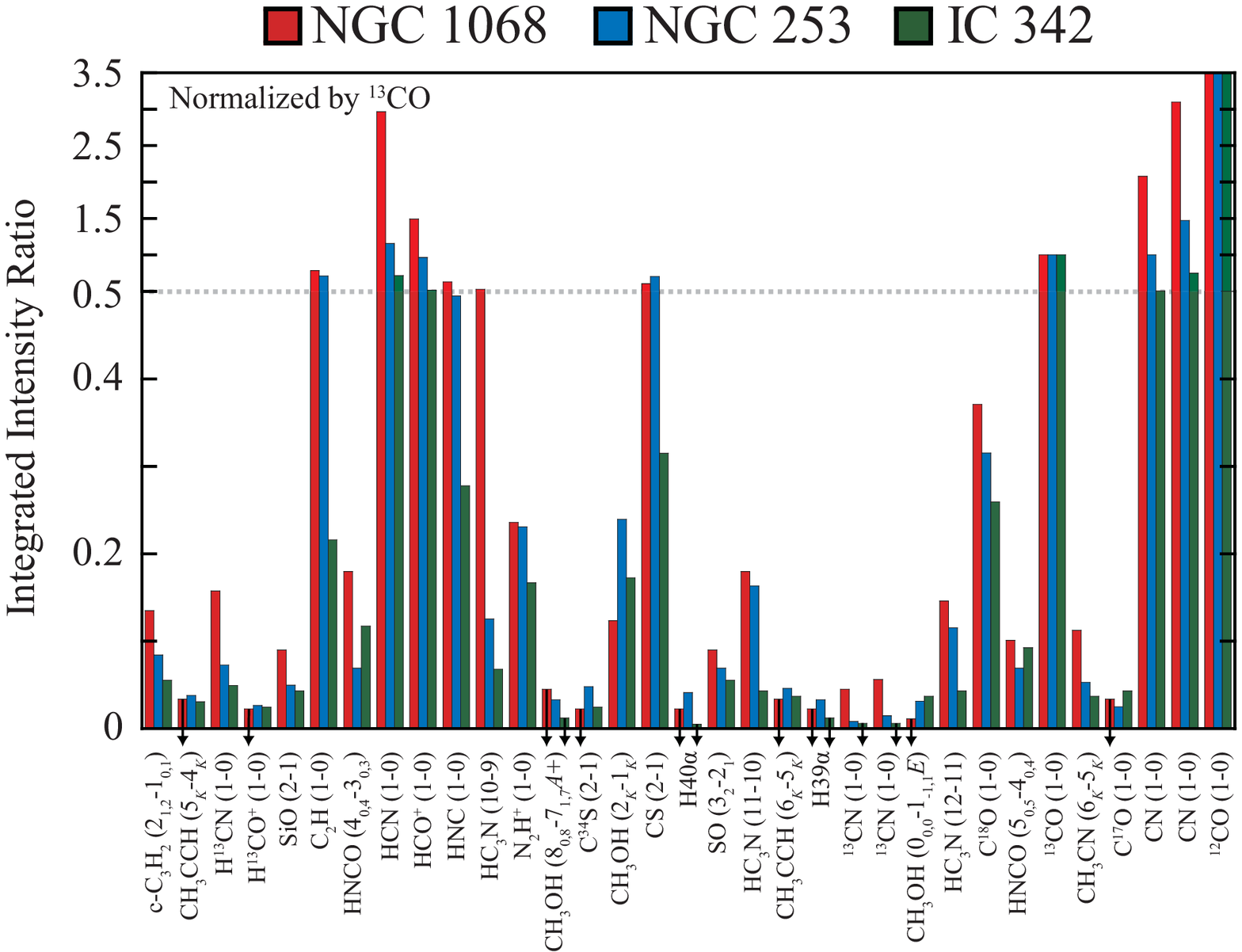} 
 \end{center}
\caption{Comparison of integrated intensity toward NGC 1068, NGC 253, and IC 342
(normalized by $^{13}$CO): The horizontal axis shows names of molecules and their
transitions
in the order of their transition frequencies. 
The vertical axis shows integrated intensity ratios in two
linear scales (0 to 0.5 below the dotted line, and 0.5 to 3.5 above
the dotted line).
The arrows indicate upper limits.}\label{fig:intint2}
\end{figure*}

The comparison of integrated intensities
among the three galaxies normalized by
that of CS or $^{13}$CO is
shown in figures \ref{fig:intint1} and \ref{fig:intint2}, respectively.
From these figures, we noticed the strongest normalized integrated 
intensities of
HCN, CN, and so on in NGC 1068.  
This characteristic is quite remarkable and is immediately
recognized also in the compressed spectra shown in figure
\ref{fig:3spectra}:
The HCN and CN lines are significantly stronger than that of $^{13}$CO
in NGC 1068.
In addition to these figures, 
selected normalized integrated intensities 
of focused molecular lines in this section 
are listed in table \ref{tab:ratio}.

\subsubsection{HCN ($J$ = 1--0) and H$^{13}$CN ($J$ = 1--0) 
intensities\label{hcnh13cn}}
The integrated intensities of HCN and H$^{13}$CN
normalized by that of CS or  $^{13}$CO are significantly 
stronger in NGC 1068 than the corresponding intensities
in NGC 253 and IC 342 as listed in table \ref{tab:ratio}.
Since the optical depth of H$^{13}$CN is much lower than
that of HCN
\citep[cf.][]{nakaji2018},
the comparison using H$^{13}$CN should be 
more reliable
for comparing the HCN intensities among the galaxies, 
and furthermore for comparing the HCN column densities.
The normalized integrated intensity of H$^{13}$CN in NGC 1068
is about 1.6--3.2 times stronger than those in NGC 253 and IC 342
(table \ref{tab:ratio}).

\subsubsection{CN ($N$ = 1--0) and $^{13}$CN ($N$ = 1--0) intensities\label{cn}}
The integrated intensities of CN 
normalized by that of CS or  $^{13}$CO are significantly 
stronger in NGC 1068 than the corresponding intensities
in NGC 253 and IC 342.
This tendency is also seen for $^{13}$CN between NGC 1068 and 
NGC 253 (non-detection in IC 342). 
Since the optical depth of $^{13}$CN is much lower than
that of CN
\citep[cf.][]{nakaji2018},
the comparison using $^{13}$CN should be 
more reliable, though the intensity is rather week.
The normalized integrated intensity of $^{13}$CN in NGC 1068
is about 4 times stronger than that in NGC 253
(table \ref{tab:ratio}).

The CN $N$ = 1--0 transition has fine and hyperfine structures.
Two lines due to the fine structure 
 ($J$ = 3/2--1/2 and 1/2--1/2) 
are mainly resolved as the broad lines of 
the external galaxies.

\subsubsection{HC$_3$N intensities\label{hc3n}}
The three transitions ($J$ = 10--9, 11--10, and 12--11) were 
detected in this survey.
The integrated intensities of HC$_3$N 
normalized by that of CS or  $^{13}$CO are relatively 
stronger in NGC 1068 than the corresponding intensities
in NGC 253 and IC 342 (table \ref{tab:ratio}).
Since the SN of the $J$ = 10--9 transition in NGC 1068
is low (2--3), the reliability of the corresponding ratios is low.

The normalized integrated intensities of the $J$ = 11--10 and 12--11
transitions in NGC 1068 are 1.1--1.5 times stronger than
those in NGC 253.
On the other hand,  they are 1.7--4.5 times stronger than 
those in IC 342.
Therefore, the characteristic of HC$_3$N is relatively weak between NGC 1068 and 
NGC 253 compared with the 
cases of HCN and CN.
We can alternatively interpret that the normalized intensities in NGC 1068
and NGC 253 are similar, and that the normalized intensities in IC 342
are relatively low.

\subsection{Characteristics of integrated intensity ratio}

\subsubsection{HCN ($J$ = 1--0)/HCO{$^+$} ($J$ = 1--0) 
integrated intensity ratios\label{r-hcnhcop}}
The  HCN/HCO{$^+$} integrated intensity 
ratios are listed in table \ref{tab:ratio-iso}.
The obtained values are 
1.98$\pm$0.11, 
1.19$\pm$0.03, and
1.38$\pm$0.02
for NGC 1068, NGC 253, and IC 342, respectively,
where the errors are  1$\sigma$.
Therefore, the ratio in NGC 1068 is 
significantly high among the three galaxies.

The normalized integrated intensity of HCN and H$^{13}$CN in NGC 1068
is significantly stronger than those in NGC 253 and IC 342 as mentioned
in section \ref{hcnh13cn} .
On the other hand,  the normalized integrated intensity of 
HCO$^+$ in NGC 1068
is also significantly stronger than those in NGC 253 and IC 342
as listed in table \ref{tab:ratio},
but it is not so strong as in the cases of HCN and H$^{13}$CN.
The  characteristic of the HCN/HCO{$^+$} integrated intensity 
ratios mentioned above is caused by this balance of the intensities. 

The H$^{13}$CN/H$^{13}$CO$^+$ integrated intensity
ratio should,
in principle,
be more reliable to study the HCN/HCO{$^+$} integrated 
intensity ratios
without the effect of the large optical depth and the 
HCN/HCO{$^+$} column density ratios, because the optical depths of the 
H$^{13}$CN and H$^{13}$CO$^+$ lines are much lower
\citep[cf.][]{nakaji2018}.
The obtained ratios are 
$>$3.1, 
2.8$\pm$0.7, and
2.0$\pm$0.6
for NGC 1068, NGC 253, and 
IC 342, respectively, as listed in table \ref{tab:ratio-iso},
where the errors are  1$\sigma$.
Therefore, the ratio in NGC 1068 is 
high among the three galaxies, but the reliability is still insufficient
due to the low SN of the lines
and to non-detection of  H$^{13}$CO$^+$ 
in NGC 1068.

\subsubsection{Integrated intensity ratios between isotopic 
species ($J$ = 1--0)\label{ratio-iso}}
The integrated intensity ratios between isotopic species, 
CO/$^{13}$CO, CO/C$^{18}$O, and HCN/H$^{13}$CN
are listed in table \ref{tab:ratio-iso}.
The ratios calculated from the data of 
NGC 1068 and
NGC 253 
obtained with the 
IRAM 30 m telescope \citep{aladro2015} 
are also listed with those obtained with the 
NRO 45 m telescope.
The ratios in NGC 1068 obtained with both of the telescopes
look significantly different each other.
On the other hand, the ratios in NGC 253 obtained with both of the 
telescopes are similar.
A possible reason is the difference in the central 
gas distributions
between NGC 1068 and NGC 253. 
The details will be discussed 
in the section \ref{d-intiso}. 

In addition, the optical depth of each line can be calculated from
the integrated intensity ratio of the isotopic species using
the 
elemental
isotopic ratios.
The results were presented in the analysis paper 
\citep{nakaji2018}.

\subsection{Comparison of integrated 
intensity in NGC 1068 between NRO 45 m and IRAM 30 m
telescopes\label{ratio-nro-iram}}
NGC 1068 has the CND and the surrounding ring-like starburst region
with the diameter of about 30$''$ as already mentioned in the introduction
with the figure \ref{fig:45mbeam}.
The observed line intensities, therefore, should sensitively reflect 
the coupling of the beams and the distributions
of the molecules (in the CND and/or the starburst region).

\begin{figure*}
 \begin{center}
    \includegraphics[width=16cm]{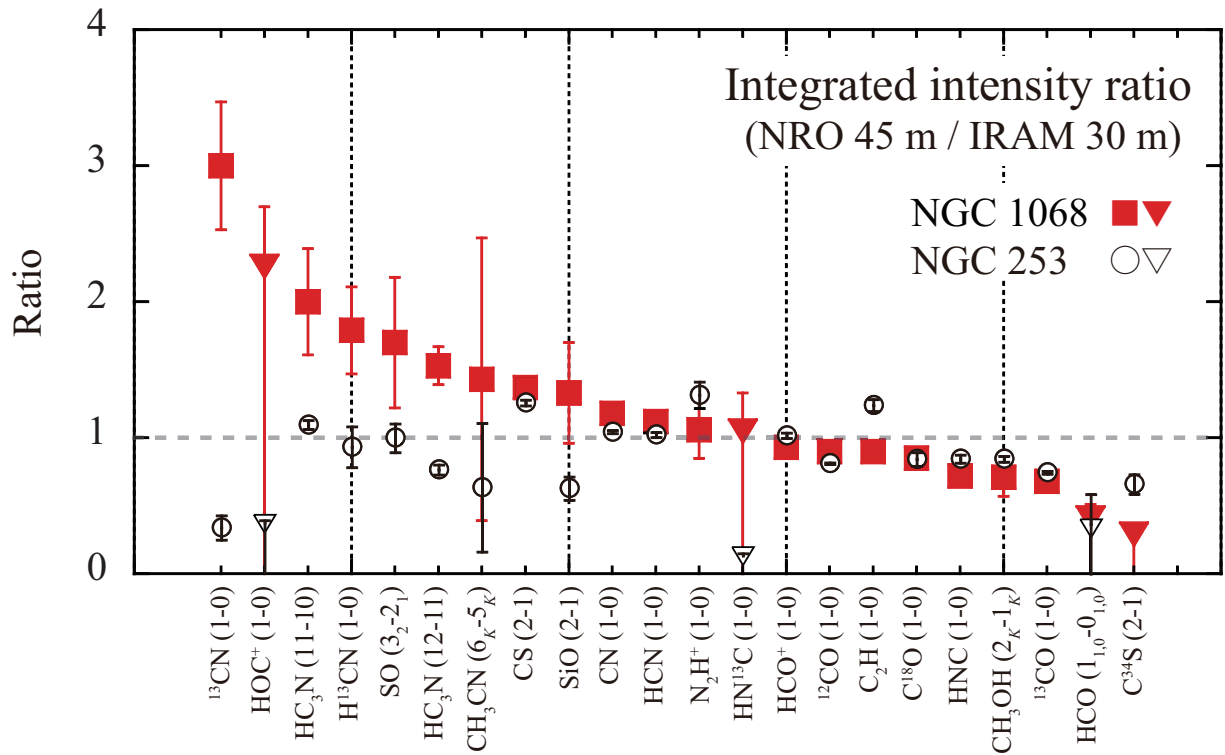} 
 \end{center}
\caption{Comparison of integrated intensity ratios 
(NRO 45 m/IRAM 30 m)
obtained with the Nobeyama 45 m and the IRAM 30 m telescopes
toward NGC 1068 (filled square and triangle) and NGC 253
(open square and triangle).
Triangles indicate upper limits.
The error was calculated based both on the 1$\sigma$ error of
the integrated intensity obtained with the 45 m telescope and on
the reported error of
the integrated intensity obtained with the 30 m telescope.
For NGC 1068 the data of $^{12}$CO and $^{13}$CN taken with the
30 m telescope were from \cite{aladro2013}, and for other
molecules from \cite{aladro2015}.
For NGC 253 the data taken with the
30 m telescope were from \cite{aladro2015}.
Since the baselines at the frequency regions of 
HOC$^+$ and HN$^{13}$C obtained toward NGC 253 with the  
45 m telescope seem to be not good,
the upper limits of HOC$^+$ and HN$^{13}$C may be underestimated. 
}\label{fig:ratio}
\end{figure*}

In this point of view, our line intensities were compared with 
those obtained with the IRAM 30 m telescope 
\citep{aladro2013, aladro2015} to estimate the 
distributions.
The results of the integrated intensity ratios 
(NRO 45 m/IRAM 30 m) are
presented in figure \ref{fig:ratio}.
In this figure, the ratios 
larger than unity
 indicate that the integrated 
intensities obtained with the 45 m telescope are stronger than the
corresponding values 
obtained with the 30 m telescope.
The ratios 
larger than unity
are for molecules
$^{13}$CN,
HC$_3$N, 
H$^{13}$CN, 
SO,
CH$_3$CN, 
CS, 
SiO,
CN,
HCN, 
and
N$_2$H$^+$
in descending order.
On the other hand, the ratios 
smaller than unity
are for molecules
HCO$^+$,
$^{12}$CO, 
C$_2$H,
C$^{18}$O,
HNC,
CH$_3$OH,
$^{13}$CO,
and
C$^{34}$S
in descending order.

The similar ratios in NGC 253 were overlaid in figure \ref{fig:ratio}
for comparison.
The ratios in NGC 253 are between $\sim$0.3--1.3 except for 
upper limits.  
This trend is significantly different from that in
NGC 1068.
The obtained ratios in both of the galaxies
and their relation to the distributions 
are discussed later
in the section \ref{d-int-ratio}. 

\section{Discussion}
\subsection{Implication for detection and non-detection\label{dis-imp}}

\begin{itemize}
\item[-] Hydrogen recombination lines (H39$\alpha$ and H40$\alpha$) are 
detected in NGC 253, but not detected in NGC 1068 and IC 342.
In NGC 253 their detections have already been reported with single-dish
telescopes and interferometers \citep[e.g.,][]{puxley1997, martin2006, 
aladro2015, bendo2015, meier2015, nakani2015}.
In NGC 1068 no lines had
been detected so far at the millimeter or 
submillimeter wavelength  \citep{puxley1991, izumi2016re}, 
but
very recently \cite{qiu2018} reported a detection of the H42$\alpha$
line at 85695.0 MHz, which is in the gap of our frequency settings, observed
with the IRAM 30 m telescope.
They suggested that the H42$\alpha$ line comes likely from 
the spiral arms considering its central velocity and the results of
 \cite{izumi2016re}.
In IC 342
recombination lines in the 5 and 6.7 GHz regions (C-band)
and in the 34 and 35 GHz regions (Ka-band) were detected recently
by averaging the spectra obtained with the Jansky Very Large Array
\citep{balser2017}.

It is interesting to note that the ionized gas in or close to the CND of NGC 1068
does not emit the radio recombination lines with detectable intensity, and 
that the HII regions in the starburst regions in NGC 253 emit
the significantly strong lines.  
The expected intensity of the lines at the submillimeter wavelength  from 
the broad line region in NGC 1068
can be detectable according to a 
model \citep{scovil2013}, 
although no detection is reported so far as shown 
by \cite{izumi2016re}.

The limited observations
 of the recombination lines in IC 342 could be
due to the relatively small amount of the total activity of the
HII regions: The H$\alpha$ emission in IC 342 is about 4.9 times
weaker than that in NGC 253 \citep{kennic2008}.

\item[-] CH$_3$CCH is often detected in starburst galaxies, but not in
galaxies with AGNs \citep[e.g.,][]{mauers1991, martin2006, aladro2011, aladro2011b, 
aladro2013, aladro2015}.
In the present study,
CH$_3$CCH 
is not detected in NGC 1068 with the relatively small
beam of the NRO 45 m telescope as single-dish telescopes, but 
it is detected in NGC 253 and IC 342.
Consequently it is possible to conclude more strictly than before 
that the CND in NGC 1068
is not in favorable condition to maintain the abundance of 
CH$_3$CCH.
Therefore, CH$_3$CCH can be a useful molecule to judge
whether the origin of activity in galaxies is starburst or AGN,
as already discussed in \cite{aladro2013, aladro2015} and 
\cite{nakaji2018}.
As mentioned in the section
\ref{impnon}, \cite{qiu2018} reported the detection of 
CH$_3$CCH ($J_K$ = 5$_0$--4$_0$) with the IRAM 30 m telescope.
In the context above, its emission may come from the starburst ring,
which is covered with the beam of the IRAM 30 m telescope.
The high spatial resolution data will be necessary to
study the distribution of CH$_3$CCH.

Recently \cite{watana2017} and \cite{nishim2017} reported 
mapping spectral line
surveys toward the Galactic active star-forming regions W51
and W3(OH), respectively.
They detected CH$_3$CCH in the active regions such as the hot core
in W51 and W3(OH), 
but its line is 
found to be missing in the averaged spectra over all the 
observed areas.
The areas are 39 pc $\times$ 47 pc for W51 and 
9.0 pc $\times$ 9.0 pc for W3(OH).
On the other hand,
our beam at 86 GHz corresponds to the
linear scales of $\sim$320 pc for NGC 253 and 
$\sim$360 pc for IC 342.
These results suggest that the starburst regions in
NGC 253 and IC 342 are not collections of the Galactic active
star-forming regions like W51 and W3(OH):
The starburst regions are suggested to be collections of 
the Galactic hot core like regions.

\end{itemize}

\subsection{Characteristics of integrated intensity (Normalized by
CS or $^{13}$CO)}
\subsubsection{HCN ($J$ = 1--0) and H$^{13}$CN ($J$ = 1--0) 
intensities\label{dis-hcn}}
In NGC 1068 HCN ($J$ = 1--0) is distributed mainly in the CND as 
already reported with the interferometric observations
including both the CND and the starburst ring in the field of
view \citep[e.g.,][]{kohno2008}
(see also section \ref{d-int-ratio}).

 The enhancement of the HCN abundance in X-ray irradiated 
regions such as the CND
is predicted by model calculations \citep[e.g.,][]{lepp1996}.
In addition, \cite{harada2010} pointed out that the HCN abundance is 
enhanced under high-temperature conditions via the hydrogenation of CN
(reaction barrier of 820 K):
 $$ \rm{CN + H_2 \rightarrow HCN + H} .$$
On the other hand, \cite{meijer2011} reported model calculations 
in extreme environments with the effects of cosmic rays and
mechanical heating (e.g., supernova driven turbulence).
According to their results, the HCN abundance does not show a 
strong response to enhanced cosmic ray rates, but the abundance
enhances with mechanical heating.

These situations are almost in accord with the proposed high-temperature
chemistry in another Seyfert galaxy NGC 1097 
based on the enhanced HCN ($J$ = 4--3) line intensity \citep{izumi2013}.
Since this galaxy has the low X-ray luminosity AGN
\citep[$L_{2-10 {\rm keV}}$ = 4.4$\times$10$^{40}$ erg s$^{-1}$,][]{nemmen2006}, 
the effect of X-rays is thought to be not efficient
\citep[cf. $L_{2-10 {\rm keV}}$ = 1$\times$10$^{43}$--10$^{44}$ erg s$^{-1}$
for NGC 1068,][]{iwasaw1997, colber2002}.

Our results of the enhancement of the normalized intensity of HCN
and H$^{13}$CN in NGC 1068 
could also be interpreted with these mechanisms above.
One of the origins of the mechanical heating in the CND in NGC 1068 can be
an AGN driven outflow, which was identified by \cite{garcia2014} based on the
analysis of the velocity field of CO ($J$ = 3--2). 

\cite{izumi2013} proposed that the intensity ratio of
HCN ($J$ = 4--3)/CS ($J$ = 7--6)  can be used to distinguish the
power sources of galaxies, AGN or starburst.
Our intensities shown in the columns in the left side in
table 5 are normalized by the intensities of 
CS ($J$ = 2--1) instead of CS ($J$ = 7--6).
As already shown, the intensity ratios of
HCN ($J$ = 1--0)/CS ($J$ = 2--1) are higher in
NGC 1068 than
those in NGC 253 and IC 342.
Therefore, the intensity ratio of HCN/CS is 
generally useful, but we should note that
the ratios obtained from the high excitation lines,
HCN ($J$ = 4--3)/CS ($J$ = 7--6), are more sensitive
to excitation conditions than those obtained from the 
low excitation lines,
HCN ($J$ = 1--0)/CS ($J$ = 2--1),
as already discussed by 
\cite{martin2015}.

\subsubsection{CN ($N$ = 1--0) and $^{13}$CN ($N$ = 1--0) intensities}

The normalized integrated intensities of CN are significantly 
stronger in NGC 1068 than the corresponding intensities
in NGC 253 and IC 342 as already reported in the section
\ref{cn}.
The enhancement of the CN abundance in X-ray irradiated 
regions is predicted by model calculations
\citep[e.g.,][]{krolik1983, lepp1996, harada2013}.
On the other hand, the CN abundance
does not enhance under high-temperature conditions 
\citep{harada2010, harada2013}.
Therefore, the present results of the strong intensity of
CN and $^{13}$CN can be mainly due to X-ray radiation.
According to the ALMA observations of
CN ($N$ = 3--2) in NGC 1068 \citep{nakaji2015},
which resolves the CND and the starburst ring,
CN is distributed only in the CND.
This result does not contradict to the formation mechanism
above.
We are analyzing the $N$ = 1--0 lines of CN in NGC 1068
obtained with ALMA,
and the results will be presented elsewhere.

\subsubsection{HC$_3$N intensities}
As explained in the section \ref{hc3n}, the normalized integrated intensity
of HC$_3$N can be relatively strong in NGC 1068.
According to the model calculations in high-temperature and/or
with AGN \citep{harada2010, harada2013},
the abundance of 
HC$_3$N is high
in the mid-plane in the CND 
shielded from X-ray radiation.
The relatively strong HC$_3$N intensity in the 
present study can be qualitatively explained with 
such models.

\cite{aladro2015} 
carried out line survey observations 
with the IRAM 30 m telescope toward
eight galaxies with starbursts, AGNs, and/or ultra-luminous
infrared emission (ULIRGs), and reported
that the abundances of HC$_3$N
are enhanced in Arp 220 and Mrk 231 in their sample of galaxies.
They mentioned that this could be related to larger amount of 
dense gas and warm dust.
\cite{costag2011} also
discussed that the emission of  HC$_3$N would be coming
from hot core-like regions
based on line intensities obtained from their line survey observations
toward 23 galaxies.
These conditions may be similar to those in the CND in NGC 1068.
On the other hand,
HC$_3$N is relatively not abundant in NGC 1068 in the observations
of \cite{aladro2015}.
The different results between the IRAM 30 m and 
the NRO 45 m telescopes 
can be due to the relatively small
beam size of the 45 m telescope, 
which observes the CND more selectively.
Actually, the HC$_3$N lines ($J$ = 11-10 and 12-11)  were found 
to be concentrated in the CND based on the ALMA
observations \citep{takano2014}.
The concentration in the CND would support the HC$_3$N formation
in the mid-plane. 

\subsection{Characteristics of integrated intensity ratio}

\subsubsection{HCN ($J$ = 1--0)/HCO{$^+$} ($J$ = 1--0) intensity ratios}
As explained in the section \ref{r-hcnhcop},
the  HCN/HCO{$^+$} integrated intensity 
ratios obtained are significantly higher in 
NGC 1068 than those in NGC 253 and IC 342.
This tendency is consistent with the results 
toward the central regions
observed with 
the Nobeyama Millimeter Array (NMA)
\citep{kohno2001}.
Our ratio in NGC 1068, 
1.98$\pm$0.11, is significantly higher than
the ratio of 1.64$\pm$0.03 obtained with the IRAM 30 m telescope
\citep{aladro2015}
(table 6), where the errors are  1$\sigma$, 
because the NRO 45 m telescope is
probing more selectively the central region, where the ratio
is high as shown below.
These ratios 
in the central region
obtained with
interferometers are
 $\sim$2.3 
\citep[NMA with the beam size of
\timeform{6.3''}$\times$\timeform{4.9''},][]{kohno2001} 
and $\sim$2.5 
\citep[Plateau de Bure interferometer with the beam size of
\timeform{6.6''}$\times$\timeform{5.1''},][]{viti2014}.

\cite{meijer2007}  reported detailed model calculations for PDR and XDR
regions. 
HCN ($J$ = 1--0)/HCO{$^+$} ($J$ = 1--0) intensity ratios are included
in their results. 
The ratios depend on
the density and the column density:
In high density ($>10^5$ cm$^{-3}$) and high column density
 ($>10^{23}$ cm$^{-2}$) conditions the ratios for the 
XDR regions are 
smaller than unity, 
which is not consistent
with our results, if the emission lines are mainly coming from
the XDR regions.
Considering the situations above, the mechanical heating mentioned in 
the section \ref{dis-hcn} can be an efficient mechanism to enhance
the HCN abundance.

In the starburst galaxy NGC 253, 
the  HCN/HCO{$^+$}
integrated intensity 
ratios obtained with the NRO 45 m and the IRAM 30 m
telescopes are similar, 1.19$\pm$0.03 and 1.18$\pm$0.02 \citep{aladro2015}, 
respectively (table 6), where the errors are  1$\sigma$.
Therefore, the situation is different from that in NGC 1068, and the 
ratio seems to be spatially rather uniform in the dense gas
in NGC 253. 
In such case, the ratio is expected to be insensitive to the 
beam sizes.
Actually, \cite{meier2015} reported that 
the HCO{$^+$}/HCN intensity ratios are $\sim$1 in the 
regions 4--10, which they defined in the central 
\timeform{35''} in NGC 253.

Another factor 
to affect the HCN/HCO{$^+$} integrated intensity 
ratios is metallicity. 
In low-metallicity galaxies such as IC 10, Large Magellanic Cloud, and M 33,
the HCN/HCO{$^+$}  integrated
intensity ratios (mainly from the $J$ = 1--0 transition) 
are 
smaller than unity
\citep[e.g.,][]{chin1997, paron2014, nishim2016a, nishim2016b, braine2017}.

In the cases of NGC 1068, NGC 253, and IC 342, 
the oxygen abundances relative to hydrogen,
12+log(O/H), are 8.87 \citep[][]{kraeme2015},
$\sim$8.5--9.0, and $\sim$8.3--9.25 
\citep[][from their figure 2]{vila1993}, respectively, 
where the solar value (solar photosphere)
is 8.66 \citep{asplun2006}.
The nitrogen abundances relative to hydrogen,
12+log(N/H), are 8.64 \citep[][]{kraeme2015},
$\sim$7.25--8.0, and $\sim$7.05--9.0 
\citep[][from their figure 2]{vila1993}
for NGC 1068, NGC 253, and IC 342, respectively, 
where the solar value (solar photosphere)
is 7.78 \citep{asplun2006}.
The values from \cite{kraeme2015} are based on
X-ray observations in the central region ($\sim$100''), and those from 
\cite{vila1993} are based on optical observations
toward giant HII regions.
The values of oxygen abundances in NGC 253 and IC 342
show scatter, but the averaged values in NGC 253 and IC 342
and the value in NGC 1068 are similar to the solar value.
Furthermore,
the values of nitrogen abundances in NGC 253 and IC 342
also show scatter, but the averaged values 
are similar to the solar value.
On the other hand, the nitrogen abundance in NGC 1068
is likely to be higher than the solar value.
In general each galaxy has gradient and scatter of metallicity.
Therefore,
the effect of metallicity viewed with the beam size of the 
NRO 45 m telescope 
may not be
significant.
Observations with high spatial resolution are necessary for 
further study of the effect of metallicity.
The details are discussed in \cite{nakaji2018}.

\cite{watana2014} carried out spectral line surveys toward the two
positions (called P1 and P2) in the spiral arm of M51 in the 3 mm and
2 mm bands.
Their results can be used as 
one of the
references for studies of molecular
abundances in  CNDs and starburst galaxies.
The HCN ($J$ = 1--0)/HCO{$^+$} ($J$ = 1--0) integrated intensity ratios are calculated
from their data to be 1.34$\pm$0.07 for the P1 position
 and 1.23$\pm$0.11 for the P2 position,
where the errors are  1$\sigma$.
These ratios are similar to those of the starburst
galaxies, 1.19$\pm$0.03 and 1.38$\pm$0.02 for 
NGC 253 and IC 342, respectively (table \ref{tab:ratio-iso}).
Thus, the ratios may not significantly depend on the 
scale of the star formation activities (starburst or not) 
in the case where there is no
significant effects of  mechanical heating.

\subsubsection{Integrated intensity ratios between isotopic species
 ($J$ = 1--0)\label{d-intiso}}
The integrated intensity ratios between isotopic species, 
are listed in table \ref{tab:ratio-iso} and already introduced in section
 \ref{ratio-iso}.
The different ratios in NGC 1068 obtained with the NRO 45 m telescope
and IRAM 30 m telescope,
for example, 17.4$\pm$0.4 and 6.2$\pm$0.1 for CO/$^{13}$CO,
where the errors are  1$\sigma$, 
indicate that the distributions of the intensities of the 
isotopic species are different each other.
Such different distributions are actually found with the high spatial
resolution data obtained with interferometers including ALMA.
The CO emission
is distributed both in the CND and the starburst ring \citep[e.g.,][]{schinn2000},
but $^{13}$CO and C$^{18}$O show weak emission lines in the CND and
mainly distributed in the starburst ring 
\citep[e.g.,][]{helfer1995, papado1996, takano2014, tosaki2017}.
Therefore, the ratios are higher with the small beam of the 45 m
telescope than those with the 30 m telescope.
The situation is contrary in the case of the ratio HCN/H$^{13}$CN.
The ratio obtained with the 45 m telescope is lower than that obtained with 
the 30 m telescope.
These ratios indicate that
the distribution of HCN is expected to have more significant 
detectable
fraction
 in the starburst ring than that of  H$^{13}$CN.
Detailed distributions of HCN and H$^{13}$CN are necessary to 
directly interpret this case.

On the other hand, the similar ratios obtained in NGC 253 with both of the 
telescopes indicate that the distributions of the emission lines of the 
species are similar each other (CO and $^{13}$CO, CO and C$^{18}$O,
HCN and H$^{13}$CN). 
Actually similar distributions between HCN and H$^{13}$CN are reported based on 
high spatial resolution observations with ALMA \citep{meier2015}.
The different situation shown above between NGC 1068 and NGC 253
is probably due to the existence of the CND, which 
significantly affects the molecular
abundance and excitation, in NGC 1068.

\subsection{Integrated intensity ratios in NGC 1068 between NRO 45 m and IRAM 30 m telescopes\label{d-int-ratio}}
The integrated intensity ratios (NRO 45 m/IRAM 30 m) 
in NGC 1068 were compared
in the section \ref{ratio-nro-iram}.
The molecules with the ratios 
larger than unity
are expected to 
have concentration mainly
in the CND, because such molecules are observed
with relatively small beam dilution with the small beam of the 
NRO 45 m telescope.
On the other hand, the molecules with the ratios 
smaller than unity
are expected to
be distributed significantly in the starburst ring.
These expectations are now becoming possible to study 
directly with high spatial resolution
data obtained with sensitive interferometers including ALMA.

The molecules with the ratios 
larger than unity
and with
their interferometric data available are actually found to 
have 
significant
concentration in the CND: 
HC$_3$N  ($J$ = 11--10, 12--11),
SO ($J_N$ = 3$_2$--2$_1$), 
CH$_3$CN ($J_K$ = 6$_K$--5$_K$) \citep{takano2014},
CS ($J$ = 2--1) \citep{taccon1997, takano2014, tosaki2017}.
SiO  ($J$ = 2--1) \citep{garcia2010},
and
HCN ($J$ = 1--0) \citep[e.g.,][]{kohno2008}.
The degree of the concentration in the CND depends on the ratio.
For example, the ratio is about 1.4 in the case of CS. 
The emission of CS is
significantly concentrated in the CND, and in addition
the emission is also clear enough to trace the starburst ring.

On the other hand, the molecules 
with the ratios 
smaller than unity
and 
with their interferometric data available
are found to be
distributed mainly in the starburst ring: 
$^{12}$CO ($J$ = 1--0) \citep[e.g.,][]{schinn2000},
$^{13}$CO ($J$ = 1--0) and C$^{18}$O ($J$ = 1--0) \citep[e.g.,][]{takano2014, tosaki2017}.
C$_2$H ($N$ = 1--0) 
and
CH$_3$OH  ($J_K$=2$_K$-1$_K$) 
also show 
the ratios 
smaller than unity, 
and 
these molecular lines are found to be 
distributed significantly both in the CND and the starburst ring
(\cite{garcia2017} for C$_2$H; 
\cite{takano2014}  and \cite{tosaki2017} for CH$_3$OH). 

\cite{nakaji2011} has already suggested that 
C$_2$H ($N$ = 1--0) is insusceptible to AGN or is
tracing cold molecular gas rather than the X-ray irradiated
hot gas
based on their observations toward NGC 1068 and NGC 253
in the early stage of this line survey.
Their suggestion  does not contradict to the discussion
above.
Furthermore,
\cite{aladro2015} has also mentioned the possibility 
that significant emission of C$_2$H ($N$ = 1--0) arises
from the star forming ring in NGC 1068 and NGC 7469
similar to the case in NGC 1097 \citep{martin2015}.

In the case of C$^{34}$S ($J$ = 2--1) the upper limit of the ratio
is much lower than unity, which is not consistent with
the ratio ($\sim$1.4) of CS ($J$ = 2--1).
Since the intensity of CS ($J$ = 2--1) in NGC 1068 is much weaker 
than those in NGC 253 and IC 342, detection of 
C$^{34}$S ($J$ = 2--1) in NGC 1068 is rather 
difficult: The detection of C$^{34}$S  ($J$ = 2--1) is tentative with 
the 30 m telescope, and it is not detected with the 45 m telescope.  
Therefore, the data with high quality are necessary to
study the integrated intensity ratio of 
C$^{34}$S in detail.

Based on the discussion above
with the interferometric data,
the estimates of the distributions using 
data obtained with the single-dish
telescopes are 
demonstrated to be possible.
In figure \ref{fig:ratio} the integrated intensity ratios
in NGC 253 were overlaid for comparison.
In contrast to the ratios in NGC 1068, the ratios 
in NGC 253 are at around unity, which means that
the variation of the ratios in NGC 253 is significantly 
smaller than that in NGC 1068.
These results in NGC 253 probably indicate that there is no
distinct structure comparable to that in NGC 1068
resolved with the beams of the NRO 45 m and/or 
the IRAM 30 m telescopes.

\subsection{General comments for IC 342}
This article reports the first high quality results of line survey observations
toward IC 342 in the 3 mm wavelength region
as mentioned in the introduction.
Generally the intensities of lines show similar trend with those of 
NGC 253 as shown in figures \ref{fig:intint1} and \ref{fig:intint2}.
This trend may be originated from the fact that 
both NGC 253 and IC 342 are starburst galaxies with no significant
indication of the AGN.
Non-detection of recombination lines in IC 342 indicates
lower activity of star-formation than that in NGC 253 as mentioned
in the section \ref{dis-imp}.
Actually the star formation rate in the central
$\sim$\timeform{10"} region in IC 342 is reported
to be $\sim$0.15 $M_\odot$yr$^{-1}$ \citep{balser2017}.
This rate is much smaller than 1.73$\pm$0.12 $M_\odot$yr$^{-1}$
in the central
\timeform{20"}$\times$\timeform{10"} region in NGC 253
\citep{bendo2015}.
The star formation rates in these galaxies were obtained 
based on the observations of recombination lines and
continuum emission, which are not affected by dust
extinction.

The linewidths of non-blended lines in IC 342 are 38--74 km s$^{-1}$,
which are significantly smaller than the corresponding widths
of 123--218 km s$^{-1}$ in NGC 253.
Note that these widths are severely affected by the SN of the lines.
As generally known, 
the important factor for the difference in the linewidths is 
the inclination angles, 31$\pm$6\degree \citep{crosth2000} 
for IC 342 (nearly face-on) 
and 78.5\degree \citep{pence1980, pence1981} for NGC 253 
(nearly edge-on).

\subsection{Future prospects}
In the present line survey observations data of molecular gas in NGC 1068,
NGC 253, and IC 342  in the 3 mm wavelength regions was
obtained with rather high spatial resolutions
as single-dish telescopes.
Higher spatial resolution images with interferometers are
essential for further study by directly resolving the central
structures such as the CND 
and the starburst ring in the case of
NGC 1068.
We have already obtained the corresponding imaging line survey
data of NGC 1068 with ALMA, 
and the results will be published in near future.

\section{Summary}
In the present line survey observations data of molecular gas in the Seyfert
galaxy NGC 1068 and the prototypical starburst galaxies
NGC 253 and IC 342 in the 3 mm wavelength regions were
obtained.
The results and discussion are summarized as follows.

\begin{enumerate}
 \item 
The observations were carried out with the Nobeyama 45 m
radio telescope and with the wide-band observing 
system, which became available
during our project.
The beam size was 15$''$.2--19$''$.1, which is rather small
among single-dish telescopes.
This beam size can mainly probe the circumnuclear disk (CND)
in NGC 1068 selectively.

\item
The numbers of lines
detected were 25, 34, and 31 for
NGC 1068, NGC 253, and IC 342, respectively.
The numbers of atomic and molecular species 
(distinguishing isotopologues) identified
were 19, 24, and 22 for
NGC 1068, NGC 253, and IC 342, respectively.
The hydrogen recombination lines were detected
only in NGC 253.

 \item
The integrated intensities normalized by
that of CS ($J$ = 2--1) or $^{13}$CO ($J$ = 1--0) were
compared among the galaxies.
As a result, the 
normalized
intensities of 
HCN (and H$^{13}$CN) ($J$ = 1--0),
CN (and $^{13}$CN) ($N$ = 1--0), and
HC$_3$N (e.g., $J$ = 11--10) in NGC 1068
were found to be stronger than those in
NGC 253 and IC 342 with our beam.
These results were discussed based on already reported
mechanisms of mechanical heating for HCN,
effect of X-rays on CN, and 
the high-temperature mid-plane
shielded from X-rays for HC$_3$N.

 \item 
The HCN ($J$ = 1--0)/HCO$^+$ ($J$ = 1--0)  integrated intensity ratios were
found to be higher in NGC 1068 than those in NGC 253 and IC 342.
The mechanical heating can be an important factor to 
affect the ratio.
Along with the non-detection of CH$_3$CCH in NGC 1068,
but detection in NGC 253 and IC 342, these molecules are
confirmed with our small beam to be good tracers to 
distinguish the power source
in galaxies.

\item 
The present integrated intensities in NGC 1068 and 
those obtained with the IRAM 30 m radio telescope by 
\cite{aladro2013, aladro2015}
in NGC 1068 were compared (Nobeyama 45 m/IRAM 30 m) 
to estimate the spatial distributions of molecules.
As a result, the above ratios were
demonstrated
to be 
useful to estimate the distributions of molecules
in the CND and/or the starburst ring.

\item
The first high quality results of line survey observations
toward IC 342 
in the 3 mm wavelength region were reported in this study.
Cyclic-C$_3$H$_2$, SO, and C$^{17}$O were 
detected for the first time.
Generally the relative intensities of lines show 
a
similar trend with those in 
NGC 253.

\end{enumerate}

\begin{ack}
We are grateful to the members of the line survey project,
which is one of the legacy projects with the Nobeyama
45 m telescope.
We thank the staff members of Nobeyama Radio Observatory 
for their support 
of the observations and for
the development of the new wide-band observing system, which
enabled us rapid survey of lines.
In particular, we thank Ryohei Kawabe for his
encouragement to this project.
We thank Hirofumi Inoue for his contribution to
the initial stage of this project.
We also thank Nanase Harada for useful comments
to the initial stage of the manuscript.
We also acknowledge Tomoka Tosaki for
sending the original reduced data for the figure 1.
This study was supported
by the MEXT Grant-in-Aid for Specially Promoted Research JP20001003. 
We used excellent databases of molecular spectroscopy: 
CDMS, JPL Catalog, NIST Recommended Rest Frequencies, 
and Splatalogue.
Data analysis was in part carried out on the open use  
computer system at the Astronomy Data Center
of the National Astronomical Observatory of Japan.
\end{ack}

\bibliographystyle{aa}
\bibliography{takano-gal}

\begin{table*} \label{tab:gal-prop}
  \tbl{Properties of the observed galaxies}{%
   \begin{tabular}{lcccclc}
      \hline
       Galaxy & $\alpha$(J2000.0) & $\delta$(J2000.0) & Distance & $V_{\rm LSR}$ & 
Morphology & Activity \\ 
                 &                         &                         & (Mpc)       &   (km s$^{-1}$)              &
                 &             \\
       \hline
       NGC 1068 &\timeform{2h42m40s.798}$^*$ & \timeform{-00D00'47".938}$^*$
 & 14.4$^{\dagger}$ &1150$^*$ & SA(rs)b$^{\ddagger}$ & AGN and circumnuclear starburst\\
       NGC 253  & \timeform{00h47m33s.3}$^{\S}$ & \timeform{-25D17'23"}$^{\S}$  
& 3.5$^{\|}$ &230 & SAB(s)c$^{\ddagger}$ &  nuclear starburst \\
       IC 342     & \timeform{03h46m48s.9}$^{\#}$ & \timeform{68D5'46.0"}$^{\#}$ 
& 3.93$^{**}$ &32$^{\dagger\dagger}$ & SAB(rs)cd$^{\ddagger}$ &  nuclear starburst \\
       \hline
     \end{tabular}}\label{tab:first}
\begin{tabnote}
$^*$\cite{schinn2000} : See also the footnote in the section of the observations.\\
$^{\dagger}$\cite{tully1988, bland1997}\\
$^{\ddagger}$\cite{devauc1991}\\
$^{\S}$\cite{martin2006}\\
$^{\|}$e.g., \cite{rekola2005, mouhci2005}\\
$^{\#}$\cite{falco1999}\\
$^{**}$\cite{tikhon2010}\\
$^{\dagger\dagger}$\cite{crosth2000}\\
\end{tabnote}
\end{table*}

\begin{longtable}{llllllll}
\caption{Parameters of the lines in NGC 1068$^*$} \label{tab:param1}\\
\hline              
Frequency$^{\dagger}$ & Molecule & Transition 
& ${T_{\rm mb}}$ & $V_{\rm LSR}$ & FWHM & $\int T_{\rm mb}dv$ 
& Comment  \\

 (MHz)        &       &           
& (mK) & (km s$^{-1}$) & (km s$^{-1}$) &  (K km s$^{-1}$)          
&             \\ 
  \hline
\endhead
\endfoot
  \multicolumn{8}{l}{\hbox to 0pt{\parbox{230mm}{
    \footnotemark[$^*$] Errors correspond to 1$\sigma$.
    \par\noindent
    \footnotemark[$^\dagger$] Frequencies are taken from \cite{lovas2004}.  In the case of a blended line due to fine and/or hyperfine structures, \\
a frequency of the strongest component is listed.
    \par\noindent
    }}}
\endlastfoot



85338.906 & cyclic-C$_3$H$_2$ & 2$_{1,2}$-1$_{0,1}$
& 5$\pm$1  & 1148$\pm$17 & 176$\pm$40 & 1.2$\pm$0.1
&  \\

 85457.299 &  CH$_3$CCH &  $J_K$=5$_K$--4$_K$
& --- & --- & --- &  $<$0.14
&  upper limit \\

86340.167 & H$^{13}$CN & $J$ = 1--0
& 7$\pm$1  & 1126$\pm$9 & 196$\pm$20 & 1.4$\pm$0.2 
&  \\

  86754.330 &  H$^{13}$CO$^+$ &  $J$ = 1--0
& ---  & --- & --- &   $<$0.15
&  upper limit \\

86847.010 & SiO & $J$ = 2--1
& 5$\pm$1  & 1092$\pm$19 & 176$\pm$46 & 0.8$\pm$0.2
& \\

87316.925 & C$_2$H & $N$ = 1--0
& 20$\pm$2    & 1117$\pm$11 & 187$\pm$24 & 4.4$\pm$0.3
& partially blended \\

 &  & $J$ = 3/2--1/2
&     &  &  & 
& with $J$ = 1/2--1/2 \\

87402.004 & C$_2$H & $N$ = 1--0
& 9$\pm$1 & 1110$\pm$31  & 313$\pm$80 & 2.6$\pm$0.3  
&  \\

 &  & $J$ = 1/2--1/2
&     &  &  & 
&  \\

87925.238 & HNCO & 4$_{0,4}$--3$_{0,3}$ 
& 6$\pm$3 & 1134$\pm$57 & 250$\pm$134 &  1.6$\pm$0.2
&  \\

88631.847 & HCN & $J$ = 1--0 
& 108$\pm$2 & 1117$\pm$3 & 233$\pm$6 & 26.4$\pm$0.6
&  \\

89188.526 & HCO$^+$ & $J$ = 1--0 
& 55$\pm$2  & 1132$\pm$5 & 236$\pm$11 & 13.3$\pm$0.7
&  \\

90663.574 & HNC & $J$ = 1--0 
& 28$\pm$2  & 1138$\pm$7 & 200$\pm$17 & 5.6$\pm$0.4
&  \\

90978.989 & HC$_3$N & $J$ = 10--9 
& 16$\pm$4  & 1078$\pm$33 & 254$\pm$77 & 4.7$\pm$1.0
&  \\

93173.777 & N$_2$H$^+$ & $J$ = 1--0
& 8$\pm$1   & 1112$\pm$16   & 245$\pm$37   & 2.1$\pm$0.4 
&  \\


96741.377 & CH$_3$OH & $J$ = 2$_K$--1$_K$ 
& 5$\pm$1  & 1104$\pm$16 & 225$\pm$40 & 1.1$\pm$0.2
& \\

97980.953 & CS & $J$ = 2--1
& 22$\pm$1  & 1123$\pm$5 & 242$\pm$12 & 5.4$\pm$0.2
&  \\

99299.905 & SO & $J_N$ = 3$_2$--2$_1$
& 4$\pm$1  & 1194$\pm$26 & 219$\pm$62 & 0.8$\pm$0.2
&  \\

100076.385 & HC$_3$N & $J$ = 11--10
& 9$\pm$2    & 1094$\pm$19 & 171$\pm$46 & 1.6$\pm$0.3
&  \\

        
108651.297 & $^{13}$CN & $N$ = 1--0 
& 2$\pm$1    & 1076$\pm$21 & 191$\pm$50 & 0.4$\pm$0.1
&  \\

 &  & $J$ = 1/2--1/2 
&  &  &  &
&  \\

108780.201 & $^{13}$CN & $N$ = 1--0 
& 3$\pm$1 & 1099$\pm$21 & 197$\pm$50 & 0.5$\pm$0.1
&   \\

 &  & $J$ = 3/2--1/2 
&  &  &  &
&  \\

109173.638 & HC$_3$N & $J$ = 12--11 
& 5$\pm$1  & 1103$\pm$16 & 229$\pm$36 & 1.3$\pm$0.1
&  \\

109782.173 & C$^{18}$O & $J$ = 1--0 
& 11$\pm$1 & 1127$\pm$10 & 320$\pm$24 & 3.3$\pm$0.2
&  \\

109905.753 & HNCO & 5$_{0,5}$--4$_{0,4}$ 
& 4$\pm$1  & 1133$\pm$15 & 209$\pm$48 & 0.9$\pm$0.1
&  \\

110201.353 & $^{13}$CO & $J$ = 1--0 
& 36$\pm$1  & 1143$\pm$5 & 241$\pm$11 & 8.9$\pm$0.2
&  \\

110383.522 & CH$_3$CN & $J_K$=6$_K$--5$_K$ 
& 3$\pm$1   & 1123$\pm$31 & 287$\pm$73 & 1.0$\pm$0.2
&  \\

112358.988 & C$^{17}$O & $J$ = 1--0
& $\sim$5   & ---   & ---   & --- 
& tentative detection \\

113191.317 & CN & $N$ = 1--0
& 63$\pm$1    & 1191$\pm$3 & 275$\pm$7 & 18.5$\pm$0.3
&  \\

 &  & $J$ = 1/2--1/2
&  &  &  & 
&  \\

113490.982 & CN & $N$ = 1--0
& 110$\pm$2  & 1119$\pm$2 & 239$\pm$5 & 27.6$\pm$0.4
&  \\

 &  & $J$ = 3/2--1/2
&   &  &  & 
&  \\

115271.202 & CO & $J$ = 1--0
& 638$\pm$16   & 1143$\pm$3 & 237$\pm$7 & 155$\pm$1
&  \\


 \hline
\end{longtable}
 

\begin{longtable}{llllllll}
\caption{Parameters of the lines in NGC 253$^*$} \label{tab:param2}\\
\hline              
Frequency$^\dagger$ & Molecule & Transition 
& ${T_{\rm mb}}$ & $V_{\rm LSR}$ & FWHM & $\int T_{\rm mb}dv$ 
& Comment  \\

 (MHz)        &       &           
& (mK) & (km s$^{-1}$) & (km s$^{-1}$) &  (K km s$^{-1}$)          
&  \\ 
  \hline
\endhead
\endfoot
  \multicolumn{8}{l}{\hbox to 0pt{\parbox{230mm}{
    \footnotemark[$^*$] Errors correspond to 1$\sigma$.
    \par\noindent
    \footnotemark[$^\dagger$] Frequencies are taken from \cite{lovas2004}.  In the case of a blended line due to fine and/or hyperfine structures, \\
a frequency of the strongest component is listed.  Frequencies of the recombination \\
lines are taken from \cite{lilley1968} and \cite{towle1996}.   
    \par\noindent
    }}}
\endlastfoot


85338.906 & cyclic-C$_3$H$_2$ & 2$_{1,2}$-1$_{0,1}$
& 33$\pm$2 & 230$\pm$6 & 155$\pm$13 & 5.1$\pm$0.4
&  \\

85457.299 & CH$_3$CCH & $J_K$=5$_K$--4$_K$
& 10$\pm$2 & 208$\pm$20 & 218$\pm$48 & 2.3$\pm$0.2 
&  \\

86340.167 & H$^{13}$CN & $J$ = 1--0 
& 24$\pm$2 & 250$\pm$8 & 184$\pm$18 & 4.4$\pm$0.7 
&  \\

86754.330 & H$^{13}$CO$^+$ & $J$ = 1--0
& 12$\pm$4  & 268$\pm$18 & 123$\pm$43 & 1.6$\pm$0.3 
& \\

86847.010 & SiO & $J$ = 2--1
& 17$\pm$2 & 227$\pm$12 & 179$\pm$28 & 3.0$\pm$0.3 
&  \\

87316.925 & C$_2$H & $N$ = 1--0
& 122$\pm$3    & 252$\pm$2 & 176$\pm$6 & 24$\pm$1 
& partially blended \\
 &  & $J$ = 3/2--1/2
&     &  &  & 
& with $J$ = 1/2--1/2  \\

87402.004 & C$_2$H & $N$ = 1--0
&  67$\pm$2   & 200$\pm$5 & 286$\pm$14 & 19$\pm$1
&  \\
 &  & $J$ = 1/2--1/2
&     &  &  & 
&  \\

87925.238 & HNCO & 4$_{0,4}$--3$_{0,3}$ 
& 23$\pm$4 & 279$\pm$11 & 172$\pm$28 & 4.2$\pm$0.7
&  \\

88631.847 & HCN & $J$ = 1--0 
& 341$\pm$7 & 268$\pm$2 & 194$\pm$4 & 70$\pm$1
&  \\

89188.526 & HCO$^+$ & $J$ = 1--0 
& 279$\pm$6 & 264$\pm$2 & 200$\pm$5 & 59$\pm$1
&  \\

90663.574 & HNC & $J$ = 1--0 
& 146$\pm$5 & 246$\pm$3 & 195$\pm$7 & 30$\pm$1
&  \\

90978.989 & HC$_3$N & $J$ = 10--9 
& 32$\pm$3  & 251$\pm$10 & 214$\pm$24 & 7.6$\pm$0.8
&  \\

93173.777 & N$_2$H$^+$ & $J$ = 1--0
& 75$\pm$5 & 237$\pm$6 & 175$\pm$13 & 14$\pm$1
&  \\

95169.516 & CH$_3$OH & 8$_{0,8}$--7$_{1,7}A+$ 
& 11$\pm$3 & 249$\pm$21 & 170$\pm$49 & 2.0$\pm$0.3 
&  \\


96412.961 & C$^{34}$S & $J$ = 2--1 
& 17$\pm$2 & 258$\pm$10 & 158$\pm$24 & 2.9$\pm$0.3    
&  \\

96741.377 & CH$_3$OH & $J$ = 2$_K$--1$_K$ 
& 70$\pm$3 & 264$\pm$5 & 202$\pm$11 & 14.5$\pm$0.3
&  \\

97980.953 & CS & $J$ = 2--1
& 208$\pm$5 & 245$\pm$2 & 197$\pm$5 & 42.8$\pm$0.4   
&  \\

99022.96 &  H atom &  H40$\alpha$  
& 10$\pm$1 & 224$\pm$15 & 244$\pm$35 & 2.5$\pm$0.2
&  \\

99299.905 & SO & $J_N$ = 3$_2$--2$_1$
& 25$\pm$2 & 247$\pm$5 & 168$\pm$12 & 4.2$\pm$0.3   
&  \\

100076.385 & HC$_3$N & $J$ = 11--10
& 55$\pm$3 & 237$\pm$5 & 176$\pm$11 & 9.9$\pm$0.2  
&  \\



102547.983 & CH$_3$CCH & $J_K$=6$_K$--5$_K$
& 22$\pm$3 & 283$\pm$9 & 144$\pm$22 & 2.8$\pm$0.5 
&  \\

106737.36 & H atom & H39$\alpha$ 
& 11$\pm$4 & 219$\pm$28 & 162$\pm$66 & 2.0$\pm$0.3 
&  \\

        
108651.297 & $^{13}$CN & $N$ = 1--0 
& 6$\pm$2 & 224$\pm$14 & 98$\pm$33 & 0.5$\pm$0.3 
&  \\

 &  & $J$ = 1/2--1/2 
&  &  &  &  
&  \\

108780.201 & $^{13}$CN & $N$ = 1--0 
& 6$\pm$1 & 234$\pm$11 & 111$\pm$23 & 0.9$\pm$0.2 
&  \\

 &  & $J$ = 3/2--1/2 
&  &  &  & 
&  \\

108893.929 & CH$_3$OH & 0$_{0,0}$--1$_{-1,1}E$ 
& 11$\pm$2 & 289$\pm$16 & 177$\pm$37 & 1.9$\pm$0.5
&  \\

109173.638 & HC$_3$N & $J$ = 12--11 
& 42$\pm$3 & 239$\pm$6 & 166$\pm$14 & 7.0$\pm$0.3
&  \\

109782.173 & C$^{18}$O & $J$ = 1--0 
& 90$\pm$4 & 268$\pm$4 & 198$\pm$9 & 19$\pm$1
&  \\

109905.753 & HNCO & 5$_{0,5}$--4$_{0,4}$ 
& 25$\pm$3 & 281$\pm$8 & 162$\pm$20 & 4.2$\pm$0.3
&  \\

110201.353 & $^{13}$CO & $J$ = 1--0 
& 299$\pm$8 & 267$\pm$2 & 196$\pm$6 & 60.6$\pm$0.4
&  \\

110383.522 & CH$_3$CN & $J_K$=6$_K$--5$_K$ 
&  16$\pm$2 & 258$\pm$12 & 189$\pm$29 & 3.2$\pm$0.2
&  \\

112358.988 & C$^{17}$O & $J$ = 1--0 
&  13$\pm$5 & 238$\pm$23 & 121$\pm$55 & 1.5$\pm$0.6  
&   \\

113191.317 & CN & $N$ = 1--0
& 243$\pm$6    & 293$\pm$3 & 236$\pm$6 & 61$\pm$1 
&  \\

 &  & $J$ = 1/2--1/2 
&  &  &  &   
&  \\

113490.982 & CN & $N$ = 1--0
& 422$\pm$9 & 247$\pm$2 & 200$\pm$5 & 89$\pm$1   
&  \\

 &  & $J$ = 3/2--1/2 
&  &  &  &     
&  \\

115271.202 & CO & $J$ = 1--0
& 4241$\pm$71    & 277$\pm$2 & 193$\pm$4 & 870$\pm$3 
&  \\

115556.253 & NS & $J$ = 5/2--3/2 ($^2\Pi_{1/2}$)
& ---    & --- & --- &  ---
&  tentative detection \\
 &  &  
&  &  &  &     
& partially blended with CO  \\


 \hline
\end{longtable}


\begin{longtable}{llllllll}
\caption{Parameters of the lines in IC 342$^*$} \label{tab:param3}\\
\hline              
Frequency$^\dagger$ & Molecule & Transition 
& ${T_{\rm mb}}$ & $V_{\rm LSR}$ & FWHM & $\int T_{\rm mb}dv$ 
& Comment  \\

 (MHz)        &       &           
& (mK) & (km s$^{-1}$) & (km s$^{-1}$) &  (K km s$^{-1}$)          
&             \\ 
  \hline
\endhead
\endfoot
  \multicolumn{8}{l}{\hbox to 0pt{\parbox{230mm}{
    \footnotemark[$^*$] Errors correspond to 1$\sigma$.
    \par\noindent
    \footnotemark[$^\dagger$] Frequencies are taken from \cite{lovas2004}.  In the case of a blended line due to fine and/or hyperfine structures, \\
a frequency of the strongest component is listed.  
    \par\noindent
    \footnotemark[$^{\ddagger}$] Five detected fine and hyperfine components of 
C$_2$H are Gaussian fitted with a common $V_{\rm LSR}$ and linewidth.\\
    The differences in frequency were fixed in the fitting.
    \par\noindent
    \footnotemark[$^{\S}$] Sum of the (partially) blended features.
    \par\noindent
    }}}
\endlastfoot


84521.206 & CH$_3$OH & 5$_{-1}$-4$_{0}$ $E$
& 7$\pm$2 & 25$\pm$9 & 67$\pm$22 & 0.5$\pm$0.1
&  \\

85338.906 & cyclic-C$_3$H$_2$ & 2$_{1,2}$-1$_{0,1}$
& 15$\pm$1 & 30$\pm$3 & 59$\pm$7 & 0.9$\pm$0.4
&  \\

85457.299 & CH$_3$CCH & $J_K$=5$_K$--4$_K$ 
& 8$\pm$1 & 48$\pm$5 & 53$\pm$11 & 0.5$\pm$0.1
&  \\

86340.167 & H$^{13}$CN & $J$ = 1--0
& 14$\pm$1 & 33$\pm$2 & 55$\pm$6 & 0.8$\pm$0.1
&  \\

86754.330 & H$^{13}$CO$^+$ & $J$ = 1--0
& 7$\pm$2 & 39$\pm$5 & 49$\pm$13 & 0.4$\pm$0.1
&  \\

86847.010 & SiO & $J$ = 2--1
 & 8$\pm$1 & 51$\pm$6 & 71$\pm$11 & 0.7$\pm$0.1
 &  \\


87316.925 & C$_2$H$^{\ddagger}$ & $N$ = 1--0
& 28$\pm$2 & 32$\pm$4 & 50$\pm$5 & 2.1$\pm$0.1$^{\S}$
& partially blended \\
 &  & $J$ = 3/2--1/2
&     &  &  & 
& with  $F$ = 1--0 \\
 &  & $F$ = 2--1
&     &  &  & 
&  \\

87328.624 & C$_2$H$^{\ddagger}$ & $N$ = 1--0
& 11$\pm$4 & 32$\pm$4 & 50$\pm$5 & ---
&  \\

 &  & $J$ = 3/2--1/2
&     &  &  & 
&\\
 &  & $F$ = 1--0
&     &  &  & 
&  \\

87402.004 & C$_2$H$^{\ddagger}$ & $N$ = 1--0
&  15$\pm$5   & 32$\pm$4 & 50$\pm$5 & 1.1$\pm$0.1$^{\S}$
& blended with $F$ = 0--1 \\
 &  & $J$ = 1/2--1/2
&     &  &  & 
&  \\
 &  & $F$ = 1--1
&     &  &  & 
&  \\

87407.165 & C$_2$H$^{\ddagger}$ & $N$ = 1--0
&  6$\pm$6   & 32$\pm$4 & 50$\pm$5 & ---
&   \\

 &  & $J$ = 1/2--1/2
&     &  &  & 
& \\
 &  & $F$ = 0--1
&     &  &  & 
&  \\

87446.512 & C$_2$H$^{\ddagger}$ & $N$ = 1--0
&  4$\pm$2   & 32$\pm$4 & 50$\pm$5 & 0.3$\pm$0.1
&  \\
 &  & $J$ = 1/2--1/2
&     &  &  & 
&  \\
 &  & $F$ = 1--0
&     &  &  & 
&  \\


87925.238 & HNCO & 4$_{0,4}$--3$_{0,3}$ 
& 37$\pm$4 & 39$\pm$3 & 49$\pm$6 & 1.9$\pm$0.1
&  \\

88631.847 & HCN & $J$ = 1--0 
& 183$\pm$3 & 33$\pm$1 & 59$\pm$1 & 11.6$\pm$0.1
&  \\

89188.526 & HCO$^+$ & $J$ = 1--0 
& 140$\pm$3 & 33$\pm$1 & 55$\pm$2 & 8.4$\pm$0.1
& \\

89487.414 & HOC$^+$ & $J$ = 1--0
& $\sim$10    & --- & --- & --- 
&  tentative detection \\

90663.574 & HNC & $J$ = 1--0 
& 82$\pm$3 & 29$\pm$1 & 50$\pm$2 & 4.5$\pm$0.1
&  \\

90978.989 & HC$_3$N & $J$ = 10--9 
& 17$\pm$3 & 41$\pm$5 & 58$\pm$12 & 1.1$\pm$0.2
&  \\

93173.777 & N$_2$H$^+$ & $J$ = 1--0
& 40$\pm$3 & 37$\pm$2 & 60$\pm$5 & 2.7$\pm$0.2
&  \\



96412.961 & C$^{34}$S & $J$ = 2--1   
& 10$\pm$1 & 35$\pm$2 & 40$\pm$6 & 0.4$\pm$0.1 
&  \\

96741.377 & CH$_3$OH & $J$ = 2$_K$--1$_K$ 
& 53$\pm$1 & 36$\pm$1  & 53$\pm$2  & 2.8$\pm$0.1 
&  \\

97980.953 & CS & $J$ = 2--1   
& 96$\pm$1 & 32.6$\pm$0.3 & 51$\pm$1 & 5.1$\pm$0.1
&  \\


99299.905 & SO & $J_N$ = 3$_2$--2$_1$   
& 13$\pm$1 & 30$\pm$2 & 60$\pm$5 & 0.9$\pm$0.1
&  \\

100076.385 & HC$_3$N & $J$ = 11--10   
& 17$\pm$2 & 37$\pm$2 & 38$\pm$5 & 0.7$\pm$0.1
&  \\



102547.983 & CH$_3$CCH & $J_K$=6$_K$--5$_K$
& 8$\pm$1 & 54$\pm$7 & 72$\pm$16 & 0.6$\pm$0.1
&  \\

108893.929 & CH$_3$OH & 0$_{0,0}$--1$_{-1,1}E$ 
& 7$\pm$1 & 35$\pm$6 & 74$\pm$15 & 0.6$\pm$0.1
&  \\

109173.638 & HC$_3$N & $J$ = 12--11 
& 13$\pm$1 & 39$\pm$3 & 53$\pm$6 & 0.7$\pm$0.1
&  \\

109782.173 & C$^{18}$O & $J$ = 1--0 
& 77$\pm$2 & 32$\pm$1 & 51$\pm$1 & 4.2$\pm$0.1
&  \\

109905.753 & HNCO & 5$_{0,5}$--4$_{0,4}$ 
& 30$\pm$2 & 40$\pm$1 & 46$\pm$3 & 1.5$\pm$0.1
&  \\

110201.353 & $^{13}$CO & $J$ = 1--0 
& 294$\pm$3 & 32.4$\pm$0.3 & 52$\pm$1 & 16.2$\pm$0.1
&  \\
110383.522 & CH$_3$CN & $J_K$=6$_K$--5$_K$ 
& 8$\pm$2 & 39$\pm$6 & 66$\pm$15 & 0.6$\pm$0.1
&  \\

112358.988 & C$^{17}$O & $J$ = 1--0
& 9$\pm$3 & 26$\pm$9 & 65$\pm$20 & 0.7$\pm$0.2
&  \\

113191.317 & CN & $N$ = 1--0 
& 23$\pm$3 & --- & 74$\pm$13 & ---
& 3 hyperfine components\\

 &  & $J$ = 1/2--1/2
& 58$\pm$2 & --- & 103$\pm$7 & 8.2$\pm$0.3$^{\S}$ 
&  are partially blended \\





113490.982 & CN & $N$ = 1--0
& 149$\pm$4 & 21$\pm$1 & 77$\pm$2 & 12.2$\pm$0.2
&  \\

 &  & $J$ = 3/2--1/2
&  &  &  & 
&  \\

115271.202 & CO & $J$ = 1--0
& 2671$\pm$18 & 32.3$\pm$0.2 & 56.3$\pm$0.4 & 160.0$\pm$0.4
&  \\


 \hline
\end{longtable}


\begin{longtable}{lllllll}
\caption{Normalized integrated intensity$^*$} \label{tab:ratio}\\
\hline              
Molecule & \multicolumn{3}{c}{Normalized by CS ($J$ = 2--1) }            & 
 \multicolumn{3}{c}{Normalized by $^{13}$CO ($J$ = 1--0) }                 \\
\cmidrule(l){2-4}
\cmidrule(l){5-7}

              &  NGC 1068           & NGC 253 & IC 342   & NGC 1068      & NGC 253 & IC 342   \\ 
\endhead
\endfoot
  \multicolumn{7}{l}{\hbox to 0pt{\parbox{230mm}{
    \footnotemark[$^*$] Errors correspond to 1$\sigma$.
    \par\noindent
    \footnotemark[$^\dagger$] Low signal to noise ratio of HC$_3$N  ($J$ = 10--9)
in NGC 1068.
    \par\noindent
    }}}
\endlastfoot
  \hline
HCN ($J$ = 1--0) &  4.9$\pm$0.2    & 1.64$\pm$0.03  & 2.27$\pm$0.05 & 
              2.97$\pm$0.09 & 1.16$\pm$0.02  & 0.72$\pm$0.01 \\

HCO$^+$ ($J$ = 1--0) &  2.5$\pm$0.2    & 1.38$\pm$0.03  & 1.65$\pm$0.04 & 
                     1.49$\pm$0.09 & 0.97$\pm$0.02  & 0.52$\pm$0.01 \\

H$^{13}$CN ($J$ = 1--0) &  0.26$\pm$0.04    & 0.10$\pm$0.02  & 0.16$\pm$0.02 & 
                     0.16$\pm$0.02 & 0.07$\pm$0.01  & 0.05$\pm$0.01 \\

CN ($N$ = 1--0, $J$ = 3/2-1/2)  & 5.1$\pm$0.2    & 2.08$\pm$0.03  & 2.39$\pm$0.06 & 
                     3.10$\pm$0.08 & 1.47$\pm$0.02  & 0.75$\pm$0.01 \\

$^{13}$CN ($N$ = 1--0, $J$ = 3/2-1/2)  & 0.09$\pm$0.02    & 0.021$\pm$0.005  & --- & 
                     0.06$\pm$0.01 & 0.015$\pm$0.003  & --- \\

HC$_3$N  ($J$ = 10-9)    &  0.9$\pm$0.2$^\dagger$ & 0.18$\pm$0.02  & 0.22$\pm$0.04 & 
                     0.5$\pm$0.1$^\dagger$ & 0.13$\pm$0.01  & 0.07$\pm$0.01 \\

HC$_3$N  ($J$ = 11-10)    &  0.30$\pm$0.06    & 0.23$\pm$0.01  & 0.14$\pm$0.02 & 
                     0.18$\pm$0.03 & 0.163$\pm$0.004  & 0.04$\pm$0.01 \\

HC$_3$N  ($J$ = 12-11)    &  0.24$\pm$0.02    & 0.16$\pm$0.01  & 0.14$\pm$0.02 & 
                     0.15$\pm$0.01 & 0.12$\pm$0.01  & 0.04$\pm$0.01 \\
\hline
\end{longtable}

\begin{longtable}{llllll}
\caption{Integrated intensity ratio of HCN/HCO$^+$ and ratios 
between isotopic species$^*$} \label{tab:ratio-iso}\\
\hline              
Ratio ($J$ = 1--0) & \multicolumn{3}{c}{NRO 45 m (Present study)}            & 
 \multicolumn{2}{c}{IRAM 30 m \citep{aladro2015}}                           \\
\cmidrule(l){2-4}
\cmidrule(l){5-6}

              &  NGC 1068           & NGC 253 & IC 342   & NGC 1068      & NGC 253   \\ 
\endhead
\endfoot
  \multicolumn{6}{l}{\hbox to 0pt{\parbox{230mm}{
    \footnotemark[$^*$] Errors correspond to 1$\sigma$. 
    \par\noindent
    \footnotemark[$^{\dagger}$] An upper limit of the H$^{13}$CO$^+$  ($J$ = 1--0)
integrated intensity is used.
See the section \ref{impnon}.
    \par\noindent
    \footnotemark[$^{\ddagger}$] H$^{13}$CO$^+$  ($J$ = 1--0)
is not detcted (or blended with SiO ($J$ = 2--1)). 
    \par\noindent
    }}}
\endlastfoot
  \hline
HCN/HCO$^+$    
&  1.98$\pm$0.11    & 1.19$\pm$0.03  & 1.38$\pm$0.02 
&  1.64$\pm$0.03 & 1.18$\pm$0.02   \\

H$^{13}$CN/H$^{13}$CO$^+$    
&  $>$3.1$^{\dagger}$    & 2.8$\pm$0.7  & 2.0$\pm$0.6 
&   ---$^{\ddagger}$ & 1.2$\pm$0.1   \\

CO/$^{13}$CO    &  17.4$\pm$0.4    & 14.4$\pm$0.1  & 9.88$\pm$0.07 & 
                                6.2$\pm$0.1 & 13.2$\pm$0.2   \\

CO/C$^{18}$O    &  47$\pm$3    & 46$\pm$2  & 38.1$\pm$0.9 & 
                                20.8$\pm$0.5 & 48$\pm$1  \\

HCN/H$^{13}$CN    & 19$\pm$3    & 16$\pm$3  & 15$\pm$2 & 
                                 30$\pm$3 & 14.5$\pm$0.4   \\
\hline
\end{longtable}

\newpage

\newpage

\newpage

\newpage

\newpage

\newpage

\newpage

\newpage

\newpage

\newpage

\newpage

\newpage

\end{document}